\documentclass[useAMS,usenatbib]{mn2e}
\usepackage{natbib}
\usepackage{graphicx}

\usepackage{subfigure}
\usepackage{amsmath}
\usepackage{amssymb}
\usepackage{longtable,lscape}
\usepackage{cite}
\usepackage{color}
\usepackage{placeins}
\usepackage{float}
\newcommand{\hi}{H{\scshape i}}
\newcommand{\hii}{H{\scshape ii}}
\newcommand{\hiir}{H{\scshape ii}~region}
\newcommand{\hiirs}{H{\scshape ii}~regions}

\title[Infrared dust bubble CS51]{Infrared dust bubble CS51 and its interaction with the surrounding interstellar medium}

\author[Das et al.]{Swagat R Das$^1$\thanks{E-mail: swagat.12@iist.ac.in}, Anandmayee Tej$^1$,  Sarita Vig$^1$, Hong-Li Liu$^{2, 3, 4}$ , Tie Liu$^5$  
\newauthor 
Ishwara Chandra C.H.$^6$, Swarna K. Ghosh$^6$ \\
$^{1}$Indian Institute Of Space Science And Technology, Trivandrum, India\\
$ ^{2}$ Department of Physics, The Chinese University of Hong Kong, Shatin, NT, Hong Kong SAR \\
$ ^{3}$ Departamento de Astronom\'ia, Universidad de Concepci\'on, Av. Esteban Iturra s/n, Distrito 
Universitario, 160-C, Chile \\
$ ^{4}$ Chinese Academy of Sciences, South America Center for Astrophysics, Camino El Observatorio 1515, Las Condes, Santiago, Chile \\
$ ^{5}$ Korea Astronomy and Space Science Institute 776, Daedeokdae-ro,
Yuseong-gu, Daejeon, Republic of Korea 305-348 \\
$ ^{6}$National Centre for Radio Astrophysics (NCRA-TIFR), Pune, India \\ }

\begin{document}

\date{}

\pagerange{\pageref{firstpage}--\pageref{lastpage}} \pubyear{}

\maketitle

\begin{center}
\begin{abstract}
A multiwavelength investigation of the southern infrared dust bubble CS51 is
presented in this paper. We probe the associated ionized, cold dust, molecular and stellar components. Radio continuum emission mapped at 610 and 1300~MHz, using the Giant Metrewave Radio Telescope, India, reveal the presence of
three compact emission components (A, B, and C) apart from large-scale diffuse
emission within the bubble interior. Radio spectral index map show the coexistence
of thermal and non-thermal emission components. Modified blackbody fits to the thermal dust emission using {\it Herschel} PACS and SPIRE data is performed to generate dust temperature and column density maps. We identify five dust clumps 
associated with CS51 with masses and radius in the range 810 -- 4600~$\rm M_{\odot}$ and 1.0 -- 1.9~pc, respectively. We further construct the column density probability distribution functions of the surrounding cold dust which display the impact of ionization feedback from high-mass stars. The estimated
dynamical and fragmentation timescales indicate the possibility of collect and collapse mechanism in play at the bubble border. Molecular line emission from the MALT90 survey is used to understand the nature of two clumps which show signatures of expansion of CS51. 

\end{abstract}

\begin{keywords}
stars: formation - ISM: HII region - ISM - radio continuum - ISM: individual objects (CS51-IRAS 17279-3350)
\end{keywords}
\end{center}

\section{Introduction}
High-mass stars ($\rm M \gtrsim 8\,M_{\odot}$) are known to have significant 
influence on the surrounding interstellar medium (ISM) given their radiative, 
mechanical and chemical feedback. The last decade has seen tremendous progress towards understanding their formation and interaction with the ISM (see reviews by \citealt {{2007ARA&A..45..481Z},{2014prpl.conf..149T}}). 
A combination of thermal pressure of the expanding \hii\ region, powerful stellar wind, and radiation pressure associated with a newly formed massive star mostly results
in a `bubble'. This manifests as a shell of enhanced density 
of swept up gas and dust between the ionization and the shock fronts encompassing a relatively low-density, evacuated cavity around the central star. The `bubbles' 
display bright-rimmed mid-infrared (MIR) morphology \citep{{1977ApJ...218..377W},{2006ApJ...649..759C},{2007ApJ...670..428C},{2008ApJ...681.1341W},{2010A&A...523A...6D},{2012A&A...542A..10A},{2012ApJ...755...71K},{2014A&A...566A..75O}}. 
The MIR emission is attributed to polycyclic aromatic hydrocarbon (PAH) molecules
in  the  photodissociation  regions  (PDRs)  surrounding O and early-B stars.
Catalogs of infrared (IR) bubbles \citep{{2006ApJ...649..759C},{2007ApJ...670..428C},{2012MNRAS.424.2442S}} show that these features are ubiquitous in 
our Galaxy. The last few years have seen a plethora of
papers on the nature of these bubbles and also on the formation of a new generation of stars in their adjacent shells or bright rimmed clumps \citep{{2007A&A...467.1125U},{2010A&A...523A...6D},{2012ApJ...756..151D},{2012ApJ...751...68L},{2013MNRAS.429.1386D},{2014MNRAS.438..426H},{2016AJ....152..152R},{2016ApJ...833...85B},{2016PASJ...68...37H},{2016ApJ...818...95L},{2017A&A...602A..95L},{2017arXiv170504907L}}.  An interesting
statistical study by \citet{2012MNRAS.421..408T} shows that the surface density of
young stellar objects (YSOs) peak towards the projected angular radius of the 
bubbles. From the association of massive YSOs with the sample of bubbles studied,
these authors further conclude that $\sim$ 14 -- 30\% of massive stars in the 
Milky Way would have possibly formed via the triggered star formation mode.
 
In this work, we focus on the southern IR bubble, CS51, which is centered at 
$\rm \alpha_{2000}=17^h31^m14.90^s$, $\rm \delta_{2000}=-33^
\circ52\arcmin55.00\arcsec$ and displays a closed ring morphology with an effective 
radius of 2.1$\arcmin$ \citep{{2007ApJ...670..428C},{2012MNRAS.424.2442S}}.
A bright IRAS source, IRAS 17279-3350, with bolometric luminosity of 1.6$\rm
\times10^5\,L_{\odot}$ \citep{2006A&A...447..221B} is found to be associated with the 
bubble. 
Several kinematic distance estimates to this  source are found in the literature. The near and far distance estimates vary between 5.1 -- 6.5~kpc and 11.7 -- 13.4~kpc, respectively \citep{{1987A&A...171..261C},{2002A&A...390.1089P},{2006A&A...447..221B},{2007ApJ...670..428C},{2013A&A...550A..21S},{2013MNRAS.431.1752U}}. In this work, we have adopted the near kinematic distance of 5.3~kpc from \citet{2007ApJ...670..428C}. The justification for adopting the near kinematic distance is the presence of \hi\ self-absorption towards the CO emission peak as
observed by \citet{1989ApJ...338..841G}(see Fig. 1 of their paper for source G354.188-0.073 which is associated with the bubble). This feature resolves the kinematic distance 
ambiguity and places objects at the near distance \citep{{2003ApJ...582..756K},{2009ApJ...699.1153R}}. 

The region associated with the bubble has been probed in radio bands by a few
authors. Radio continuum observations at 4.8 and 8.6~GHz by \citet{2003A&A...407..957M} reveal the presence of three components (named A, B, and C), with A being associated with IRAS point source. These
components are also studied at 5-GHz \citep{1994ApJS...91..347B} and at 1.4~GHz \citep{1990ApJS...74..181Z}. From the 1.2-mm observation, using the SEST 15-m telescope, \citet{2006A&A...447..221B} identified two massive dust clumps of masses 665 and 764~$\rm M_{\odot}$. The peak positions of these clumps correlate with that of the components A and C, respectively.
Non-detection of $\rm H_2O$, OH and $\rm CH_3OH$ masers towards IRAS 17279-3350 was reported in 
\citet{1996A&AS..115..285C} and references therein. 
However, in a later survey, the 6.7-GHz methanol maser was detected towards component C \citep{2010MNRAS.404.1029C}.

\begin{figure*}
\centering
\includegraphics[scale=0.3]{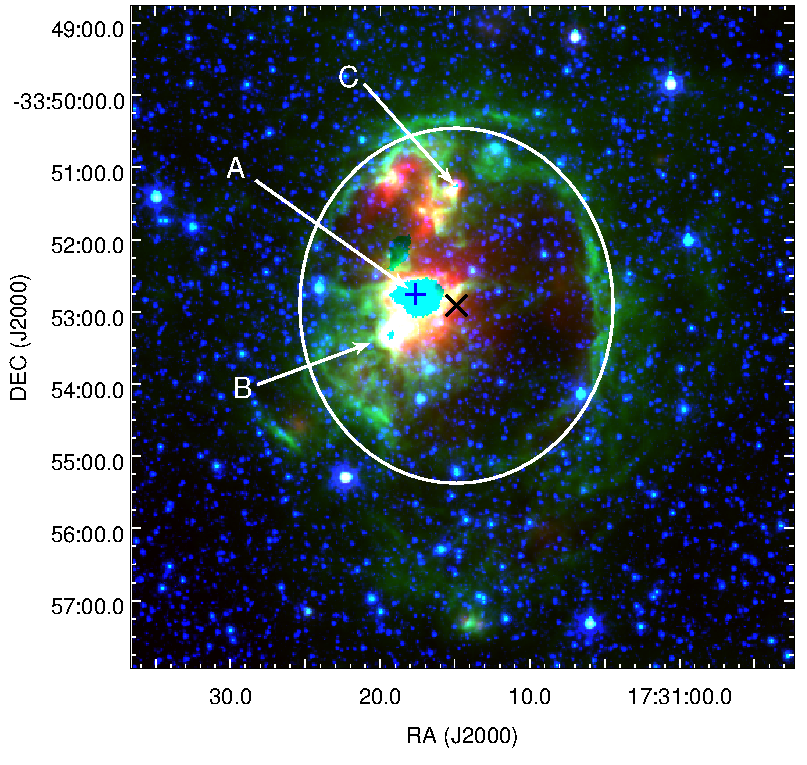}
\includegraphics[scale=0.3]{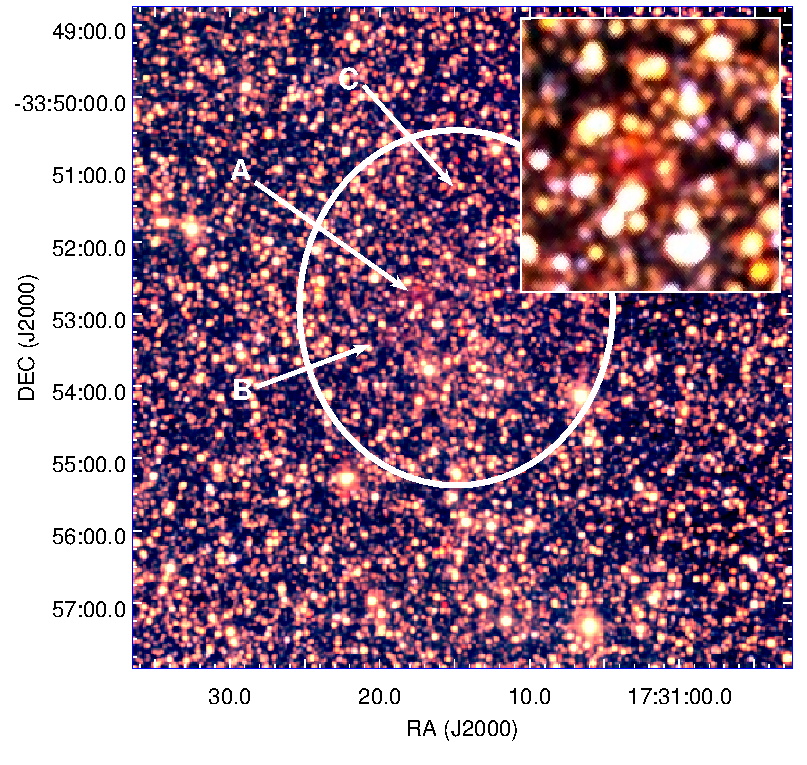}
\caption{IR colour-composite images of the region towards CS51. Left: 
The IRAC-MIPSGAL MIR colour-composite image
with 24~$\rm \mu m$ (red), 8~$\rm \mu m$ (green) and 4.5~$\rm \mu m$ (blue). 
The cross mark ($\times$) shows the position of the bubble center \citep{{2007ApJ...670..428C},{2012MNRAS.424.2442S}}. The position of IRAS 17279-3350 associated with
the bubble is shown as a `+' symbol in blue. The 24-$\rm \mu m$ emission is
saturated at this location. 
The locations of the three identified components are also highlighted with arrows where the arrow heads are the positions of the
components \citep{2003A&A...407..957M}. The ellipse shows the extent of the bubble as described \citet{2007ApJ...670..428C}. 
Right: 2MASS NIR colour-composite image with $\rm K_s$ band (red),  H band (green) and J band (blue). The inset shows the faint nebulosity seen towards A.}
\label{rgb_24_8_45}%
\end{figure*}

Figure \ref{rgb_24_8_45} shows the MIR and near-infrared (NIR) view of the region associated with CS51. The displayed morphology in the MIR is similar to those observed in other bubbles \citep{{2008ApJ...681.1341W},{2012A&A...544A..39J},{2014A&A...565A...6S},{2016ApJ...818...95L}}. 24-$\rm \mu m$ emission
sampling the hot dust is enclosed within the bright-rimmed 8-$\rm \mu m$ emission
 which shows a broken morphology towards the south. The 
8-$\rm \mu m$ emission is seen to be extended in the form of an envelope beyond the identified bubble boundary. However, the NIR colour-composite image does not reveal any bubble signature and the region is seen to be densely populated. A faint, compact $\rm K_s$-band nebulosity is seen towards IRAS 17279-3350 (see inset).

In presenting our study on the GLIMPSE bubble CS51, we have organized the paper in the following way. Section \ref{data_obs_archive} deals with the radio continuum observation and associated data reduction procedure. In this section, we have also discussed the details of various archival data used in this study. In Section \ref{result}, we discuss the results obtained using various datasets. Section \ref{feedback} focuses on the feedback of the high-mass star responsible for
the formation of CS51 and its impact on the surrounding ISM and in Section \ref{summary}, we summarize our results.

\section{Observations and archival data sets}\label{data_obs_archive}
\subsection{Radio Continuum Observations} 
\label{radio_gmrt}
To probe the ionized gas associated with the bubble, we have carried out low-frequency radio continuum observations at 610 and 1300~MHz with a bandwidth of
32~MHz using Giant Metrewave Radio Telescope (GMRT), Pune India. GMRT has a hybrid configuration with 30 fully steerable antennae each of 45-m diameter placed in a `Y' shaped array (for details about GMRT, see \citealt{1991CuSc...60...95S}). The central square of GMRT contains 12 antennae randomly distributed within 1$\times$1 $\rm km^2$, with the lowest baseline, $\sim$100~m, sensitive to diffuse emission. The remaining 18 antennae are located on the arms with 6 on each arm. This renders the array capable of producing high angular resolution with the largest baseline of $\sim$ 25~km. 

The radio observations were carried out in the spectral line mode to minimize the effects of bandwidth smearing and narrow-band RFI.
Radio sources 3C48 and 3C286 were used as primary flux calibrators and 1714-252 was used as phase calibrator for estimating the amplitude and phase gains for flux and phase calibration of the measured visibilities.
The continuum data reduction is carried out using the Astronomical Image Processing Software (AIPS). The data sets are carefully scrutinized to identify bad data (dead antenna, bad baselines, RFI, spikes, etc.) using the tasks {\tt UVPLT} and {\tt TVFLG}. Subsequent editing of these are carried out using the tasks {\tt TVFLG} and {\tt UVFLG}. To keep the bandwidth smearing effect negligible the data sets are averaged in frequency using the task {\tt SPLAT}. Low-frequency radio maps are generated by Fourier inversion and subsequent cleaning using the task {\tt IMAGR}, in which we set the `robustness' parameter to +1 (on a scale, where +4 represents nearly natural weighting and -4 is close to uniform weighting of the baselines). Wide-field imaging technique is used in order to account for the $w-$term effect (non-coplanarity). Several iterations of `phase-only' and a final iteration of `amplitude and phase' self calibration are performed to minimize the amplitude and phase errors and to reduce the $rms$ noise in the maps. The final 
maps are then primary beam corrected using the task {\tt PBCOR}. 

Galactic diffuse emission contributes to the rise in system temperature, while observing close to the Galactic plane. At the low frequencies of our radio
observations (especially at 610-MHz), the contribution from the Galactic diffuse emission becomes significant. Since the flux calibration is done using sources 
off the Galactic plane, appropriate scaling of flux densities in the final images is essential. 
The scaling factor is estimated under the assumption that the Galactic diffuse emission follows a power-law spectrum and  $T_{sky}$ at frequency $\nu$ for the target position is determined using the following equation
\begin{equation}
T_{sky}=T_{sky}^{408} \left (\frac{\nu}{\rm 408~MHz}\right )^{\gamma}
\end{equation}
where, $\gamma$ is the spectral index of the Galactic diffuse emission and is taken as -2.55 \citep{1999A&AS..137....7R}. Here, the sky temperature, $T_{sky}$, is estimated by using the measurements obtained from the all-sky 408-MHz survey of \citet{1982A&AS...47....1H}. This yields scaling factors of 2.26 and 1.27 for 610 and 1300~MHz, respectively.
Details of the GMRT radio continuum observations and maps are given in Table \ref{radio_obs}.

\begin{table}
\caption{Details of the radio interferometric continuum observations towards CS51.} 
\label{radio_obs}
\begin{tabular}{lcc}
\\
\hline \hline
Details & 610~MHz & 1300~MHz \\
\hline
Date of Obs. & 08 August 2014 & 14 August 2014 \\
Flux Calibrators & 3C286,3C48 & 3C286,3C48\\
Phase Calibrators & 1714-252 & 1714-252\\
Integration time & $\sim$~5~hr & $\sim$~5~hr \\
Synth. beam & 11.5\arcsec$\times$9.9\arcsec & 9.9\arcsec$\times$6.5\arcsec \\
Position angle. (deg) & 29.8 & 64.7 \\
{\it rms} noise (mJy/beam) & 0.5 & 0.3 \\
\hline
\end{tabular}
\end{table}

\subsection{Archival data sets}
\subsubsection{Near-infrared data from 2MASS}\label{2mass_archive}
NIR ($\rm JHK_s$) photometric data for point sources associated with CS51 are obtained from the Two Micron All Sky Survey (2MASS)\footnote{This publication makes use of data products from the Two Micron All Sky Survey , which is a joint project of the University of Massachusetts and the Infrared Processing and Analysis Center/California Institute of Technology, funded by the NASA and the NSF.} \citep {2006AJ....131.1163S} Point Source Catalog. The 2MASS images have resolution of $\sim$ 5.0$\arcsec$. Source selection is based on the ``read-flag" that gives the uncertainty in the magnitude. In our sample, we retain only those sources with ``read-flag" value of 2. The 2MASS data are used to study the nature of the stellar population which are likely to be associated with the bubble. 

\subsubsection{Mid-infrared data from Spitzer} \label{spitzer_archive}
MIR data of the region towards CS51 have been obtained from the 
archives of the {\it Spitzer} Space Telescope. Photometric data and 
images at wavelengths of 3.6, 4.5, 5.8, 8.0~$\rm \mu m$ have been retrieved
from the Galactic Legacy Infrared Midplane Survey Extraordinaire (GLIMPSE; \citealt{2003PASP..115..953B}). These images have resolution of $ < 2\arcsec$. 
 We have obtained photometric data in IRAC bands from the `highly reliable' catalog of GLIMPSE II Spring'08 catalog. 
The 24-$\rm \mu m$ image is obtained from the MIPSGAL survey \citep{2004ApJS..154...25R} and has a resolution of $\sim 6\arcsec$. 24-$\rm \mu m$ photometric data is extracted from the MIPSGAL catalog \citep{2015AJ....149...64G}.
Using these data, the structure of the bubble and distribution of YSOs are investigated. 

\subsubsection{Far-infrared data from Herschel} \label{herschel_archive}
Far-infrared (FIR) data for the bubble are obtained from the {\it Herschel} Space Observatory archives. The images obtained were observed as part of the Herschel infrared Galactic plane Survey (HI-GAL; \citealt{2010A&A...518L.100M}). We have used the images at wavelengths $\rm 70 - 500~\mu m$ obtained from the Photodetector Array Camera and Spectrometer (PACS; \citealt{2010A&A...518L...2P}) and Spectral and Photometric Imaging Receiver (SPIRE; \citealt{2010A&A...518L...3G}). The images have resolutions of 5, 11.4, 17.9, 25 and 35.7$\arcsec$ at 70, 160, 250, 350, and 500~$\rm \mu m$, respectively. We have used the FIR data to study the physical properties of cold dust emission associated with the bubble.

\subsubsection{Molecular line data from MALT90 survey}
MALT90 (Millimeter Astronomy Legacy Team 90 GHz) survey was aimed in part at characterizing the molecular clumps in the vicinity of \hiirs.\ The survey was carried out using the ATNF Mopra 22-m telescope with simultaneous mapping of 16 molecular lines near 90~GHz with a spectral resolution of 0.11~$\rm km\ s^{-1}$. The source list was taken from the ATLASGAL\footnote{ ATLASGAL is a collaboration between the Max Planck Gesellschaft (MPG: Max Planck Institute for Radioastronomy, Bonn and the Max Planck Institute for Astronomy, Heidelberg), the European Southern Observatory (ESO) and the University of Chile.} 870-$\rm \mu m$
continuum survey which detects both cold and warm clumps. We have retrieved the data from the MALT90 site (http://malt90.bu.edu). The details regarding the survey can be found in \citet{2011ApJS..197...25F}, \citet{2013PASA...30...38F} and \citet{2013PASA...30...57J}. Data reduction was performed using GILDAS (Grenoble Image and Line Data Analysis Software)\footnote{https://www.iram.fr/IRAMFR/GILDAS/}.
We mainly explore the physical properties of the cold dust clumps using this data.

\section{Results and Discussion} \label{result}
\subsection{Ionized emission}\label{radio_emission}
\begin{figure*}
\centering
\includegraphics[scale=0.3]{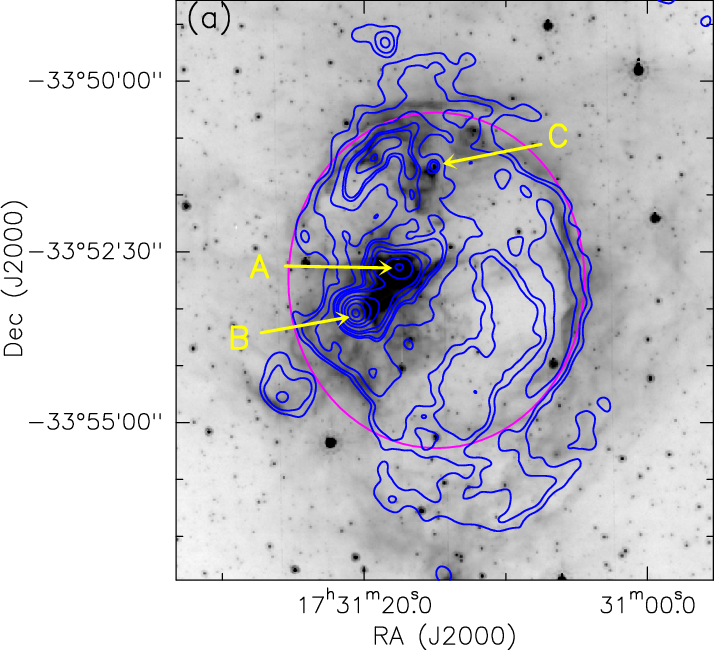}
\hspace{1cm}
\includegraphics[scale=0.3]{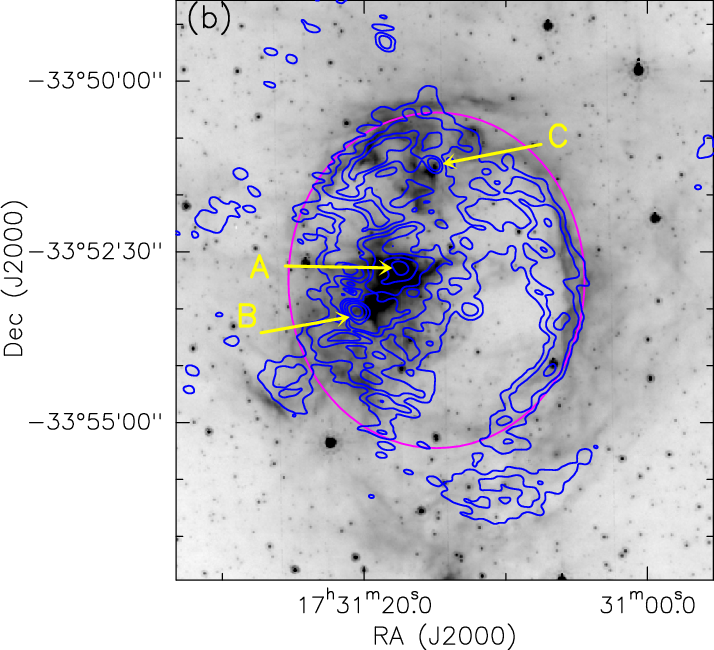}
\caption{Ionized emission associated with the bubble CS51. 
Left: Radio continuum emission at 610~MHz associated with CS51 shown as
contours overlaid on the 8-$\rm \mu m$ IRAC image. 
Right: Same as the left figure for the 1300-MHz emission.
In both the figures, the contour levels are 0.3, 1.0, 2.5, 5, 12, 25, 40, 60,
and 90\% of the peak values. The peak fluxes at 610 and 1300~MHz are 0.34 and 0.38~Jy/beam, respectively.
The positions of the identified compact components A, B, and C
(see text) are marked. The magenta ellipse shows the extent of the bubble as described in \citet{2007ApJ...670..428C}.}
\label{cs51_ionized}%
\end{figure*}

Radio emission, mapped at 610 and 1300~MHz, probing the ionized gas associated with 
CS51 is shown in Figure \ref{cs51_ionized}. Emission at both frequencies display 
complex morphology. 
Apart from an elongated cavity (a void in emission) seen towards the south-west, the bubble
interior is filled with diffuse emission.
The cavity is much more 
pronounced in the 1300~MHz map. This gives the radio emission a broken shell-like morphology.
Faint, diffuse emission extends beyond the identified bubble rim especially
towards the south. The maps also reveal three compact regions of enhanced
emission. Out of these, one is closer to the 
likely centre of the bubble, the second (and brightest of them) is more towards
the south-east periphery and the third one is located in the north-west rim of the 
bubble. The peak positions of these components at 610 and 1300~MHz coincide within 
$\sim 4\arcsec$. 

The overall structure described above is consistent with the 4.8 and 8.6-GHz maps
obtained using the Australian Telescope Compact Array (ATCA) and presented in
\citet{2003A&A...407..957M}. Their 4.8-GHz map shows patchy, diffuse emission
extending in the north-south direction. Spatial correlation with the MIR images 
clearly shows that most of this emission lies within the confines of the bubble. 
As mentioned in the introduction, \citet{2003A&A...407..957M} have also identified three distinct components namely A, 
B, and C with A being extended and B and C as compact. The components B and C
appear detached from the main complex A. The locations of these
components agree excellently (within 2.5\arcsec~from the peak positions in the 1300-MHz map) with the peaks of the detected compact components in the GMRT maps and hence
we retain the same nomenclature. Their
higher resolution 8.6-GHz map probes only the core of component A as most of the
diffuse emission is resolved out. The flux densities and the sizes of the bubble
as well as the three components determined from the GMRT and ATCA maps are listed in
Table \ref{radio_flux}. To estimate the flux and size of the components, we use the 
following procedure outlined in \citet{2016AJ....152..146N}. We first determine the threshold above which the diffuse 
emission becomes significant. Starting from the peak flux density, we sequentially
plot contours as a percentage of the peak flux density and go upto the level 
where the contribution from the diffuse emission becomes apparent. This occurs
at 40, 20, and 50\% of the peak flux density for the components A, B, and C, respectively. 
Next, this threshold emission level is given as input in the
2D {\it clumpfind} code of \citet{1994ApJ...428..693W} to determine the sizes and
integrated flux densities of the components which are considered as individual
clumps in this algorithm. The size is taken as the geometric mean of the full width half maximum ($FWHM$) for x-axis, $FWHM_x$ and for y-axis, $FWHM_y$. 
It should be noted here that completely decoupling the contribution of diffuse emission is difficult. Except for a larger (factor of $\sim 3$) peak flux density for component A,
our estimated values of peak and integrated flux densities of the three components at 1300~MHz are consistent with that obtained by \citet{1990ApJS...74..181Z} at 1400~MHz.
\begin{table*}
\tiny
\caption{GMRT and ATCA \citep{2003A&A...407..957M} results. The peak coordinates, peak and integrated flux densities, and sizes of CS51 and the individual components
are listed.}
\label{radio_flux} 
\begin{tabular}{cccccccccccc}
\\ \hline
 & \multicolumn{2}{c}{\underline{~~~~~~~~~~Peak Coordinates~~~~}} & \underline{~~Radius~~} &\multicolumn {4}{c}{\underline {~~~~~~~~~~~~~~~~Peak flux (Jy/beam)~~~~~~~~~~~~~~~}} & \multicolumn{4}{c}{\underline{~~~~~~~~~~~~~~~~~~~~~~Integrated flux (Jy)~~~~~~~~~~~~~~~~~~~~~~}}   \\
& RA (J2000)  & DEC (J2000) & (arcsec) & 610~MHz & 1300~MHz & 4.8~GHz & 8.6~GHz & 610~MHz$^\ddagger$ & 1300~MHz$^\ddagger$ & 4.8~GHz & 8.6~GHz \\
& $(^h~~^m~~~^s)$ & (~$^\circ~~\arcmin~~~\arcsec$) &  & &  & &  & & & & \\
\hline 
CS51 & 17 31 20.58 & -33 53 24.39 & 126$^{\dagger}$ & 0.34 & 0.38 & -- & -- &  5.62$\pm$0.56 & 4.07$\pm$0.40 & -- & -- \\
(entire bubble) & & & & & & & \\
\hline
\multicolumn{12}{c}{Components} \\
\hline
A & 17 31 17.53 & -33 52 43.39 & 7.3$^{\ast}$ & 0.15 & 0.13 & 0.02 & 0.01 & 0.35$\pm$0.03 & 0.45$\pm$0.04 & -- & 0.52 \\
B & 17 31 20.58 & -33 53 24.39 & 4.8$^{\ast}$ & 0.34 & 0.38 & 0.09 & 0.05 & 0.96$\pm$0.09 & 0.51$\pm$0.05 & 0.10 & 0.06 \\
C & 17 31 15.04 & -33 51 14.40 & 4.1$^{\ast}$ & 0.02 & 0.03 & 0.08 & 0.05 & 0.03$\pm$0.003 & 0.04$\pm$0.004 & 0.12 & 0.11  \\
\hline
\end{tabular} 
\\ $^{\dagger}$ Effective radius from \citet{2007ApJ...670..428C}, 
$^{\ast}$ Estimated from 2D $\it clumpfind$ algorithm output,
$^\ddagger$ Error in integrated flux has been calculated following the equation from \citet{2013ApJ...766..114S} $\rm [(2\sigma(\theta_{source}/\theta_{beam})^{1/2})^2 + (2\sigma_{flux-scale})^2 ]^{1/2}$, where $\sigma$ is the {\it rms} noise level of the map, $\rm \theta_{source}$ and $\rm \theta_{beam}$ are the size of the source and the beam, respectively, and $\rm \sigma_{flux-scale}$ is the error in the flux scale, which takes into account the uncertainty on the calibration applied to the integrated flux of the source. For GMRT maps, uncertainty in the flux calibration is taken to be 5\% \citep{2007MNRAS.374.1085L}.
\end{table*}

We derive various physical parameters associated with the ionized emission of
the bubble. 
Assuming the radio emission at 1300~MHz to be optically thin and emanating from a homogeneous, isothermal medium, we
derive the Lyman continuum flux ($\rm N_{lyc}$) required to maintain ionization in 
the nebula, the number density of electrons ($\rm n_e$) and the emission measure (EM) using the following expressions \citep{2016A&A...588A.143S}

\begin{equation}
\begin{split}
\rm \left( \frac{N_{Lyc}}{\rm sec^{-1}}\right) = 4.771 \times 10^{42} \left(\frac{\rm S_\nu}{\rm Jy}\right) \left( \frac{T_e}{\rm K}\right)^{-0.45} \\
\rm \hspace{3cm} \times \left( \frac{\nu}{\rm GHz}\right)^{0.1} \left( \frac{\rm D}{\rm pc}\right) ^{2}
\end{split}
\end{equation}

\begin{equation}
\begin{split}
\rm \left( \frac{n_e}{\rm cm^{-3}}\right)  = 2.576 \times 10^6 \left(\frac{\rm S_\nu}{\rm Jy}\right)^{0.5} \left( \frac{T_e}{\rm K}\right)^{0.175} \\
\rm   \hspace{-1cm}  \times \left( \frac{\nu}{\rm GHz}\right)^{0.05} \left( \frac{\theta_{\rm source}}{\rm arcsec}\right)^{-1.5} \left( \frac{\rm D}{\rm pc}\right) ^{-0.5}
\end{split}
\end{equation}

\begin{equation}
\begin{split}
\rm \left( \frac{\rm EM}{\rm pc\ cm^{-6}}\right) = 3.217 \times 10^7  \left(\frac{\rm S_\nu}{\rm Jy}\right) \left( \frac{T_e}{\rm K}\right)^{0.35} \\
\rm \hspace{3cm} \times \left( \frac{\nu}{\rm GHz}\right)^{0.1} \left( \frac{\theta_{\rm source}}{\rm arcsec}\right)^{-2}
\end{split}
\end{equation}
where, $\rm N_{lyc}$ is the Lyman continuum photons per second, $\rm T_e$ is the electron temperature, $\rm \nu$ is the frequency, $\rm S_{\nu}$ is the integrated flux, D is the kinematic distance to the bubble, $\theta$ 
refers to the angular size of the region considered (the entire bubble
or the components). 
We determine the electron temperature adopting the following expression \citep{2006ApJ...653.1226Q} 
\begin{equation}
\rm T_e = \left(5780 \pm 350 \right) + \left(287 \pm 46 \right) R_G
\end{equation}
From the estimated Galactocentric distance ($R_G$) of 3.3~kpc to CS51, we determine the electron temperature to be $6700 \pm 380$~K. This is higher than the value 
of 5300~K quoted in \citet{1987A&A...171..261C} derived using radio recombination
line measurements. These authors derived the electron temperature for the radio source, which is located at an angular distance of $\sim$43$\arcsec$ from the IRAS source. For further analysis, we use the electron temperature estimate of $6700 \pm 380$~K. Table \ref{radio_param} lists the derived radio properties of the bubble
as well as the three components. The table also lists results from \citet{2003A&A...407..957M}.

The estimated Lyman continuum photon flux of $\rm 1.1 \times 10^{49} s^{-1}$ 
indicates that atleast one O6V - O5.5V star or a cluster of less massive stars
are required to produce this flux (using Table 1 of \citealt{2005A&A...436.1049M}) 
and maintain ionization in the nebula associated with the bubble CS51. 
This estimate is with
the assumption of optically thin emission and hence serves as a lower limit as the emission could be optically thick at 1300 MHz. Various studies in the literature have also shown that dust absorption
of Lyman continuum photons can be very high \citep{{2001AJ....122.1788I},{2004ApJ...608..282A},{2011A&A...525A.132P}}.
Assuming a uniform distribution and
taking the derived value of electron density as a representative average value
for the entire bubble, we estimate the mass of ionized gas ($\rm M_{ion} = \frac{4}{3} \pi r_{Hii}^3 n_e m_p$, where $\rm r_{Hii}$ is taken to be the radius of the bubble and $\rm m_p$ is mass of proton) to be $\sim$300~$\rm M_{\odot}$. However, it should be 
noted that some ionized gas is seen beyond the defined bubble radius and its
contribution has not been taken into account.

The components A, B, and C could be harbouring massive stars and thus internally ionized. The derived Lyman continuum photon flux values suggest spectral types of  O9 - O8.5, O8.5 - O8, and B0.5 - B0 for A, B, and C, respectively. Component C
is more likely internally heated as there exists a 70~$\rm \mu m$ point source  \citep{2017arXiv170505693M}
coinciding with the peak position. However, for components A and B we cannot
rule out the possibility of these being externally ionized clumps due to
density inhomogeneities. Such compact ionized objects are seen in several extended \hii\ regions \citep{{1998ApJ...492..635G},{2003ApJ...596..362K},{2001ApJ...549..979K}}.

\begin{table*}
\centering
\footnotesize
\caption{Physical parameters of the bubble and the three components. The values given in parenthesis are from \citet{2003A&A...407..957M}.}
\label{radio_param} 
\begin{tabular}{ccccc}
\\ \hline
 & $\rm log\ N_{lyc}$ & $\rm n_e$ & EM & Spectral Type   \\
& ($\rm sec^{-1}$) & ($\rm \times\ 10^3\ cm^{-3}$) & ($\rm \times\ 10^6\ pc\ cm^{-6}$) &  \\
\hline
CS51 & 49.03 & 0.08 & 0.04  & O6V - O5.5V \\
(entire bubble) & & &  &\\
\hline
\multicolumn{5}{c}{Components} \\
\hline
A & 48.07 (47.88) & 2.00 (1.10) & 1.50 (0.6) &  O9 - O8.5\\
B & 48.12 (47.15) & 3.94 (9.70) & 3.89 (6.3) &  O8.5 - O8\\
C & 47.02 (47.23) & 1.42 (7.60) & 0.43 (4.8) & B0.5 - B0$^a$\\
\hline
\end{tabular} 
\\ $^a$ spectral type obtained using Table II from \citet{1973AJ.....78..929P}
\end{table*}

\subsection{Spectral index maps}\label{spectralindex}
\begin{figure*}
\centering
\includegraphics[scale=0.3]{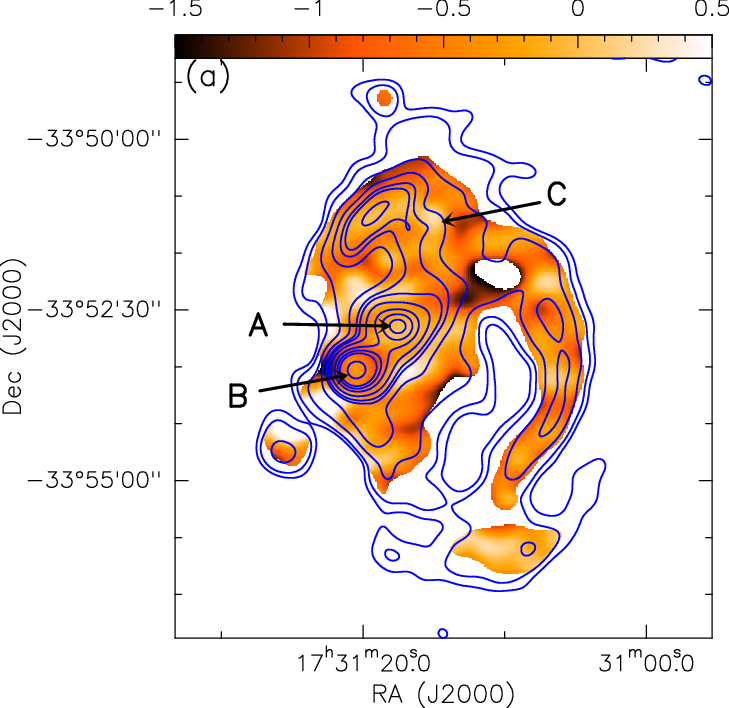}
\hspace{1cm}
\includegraphics[scale=0.3]{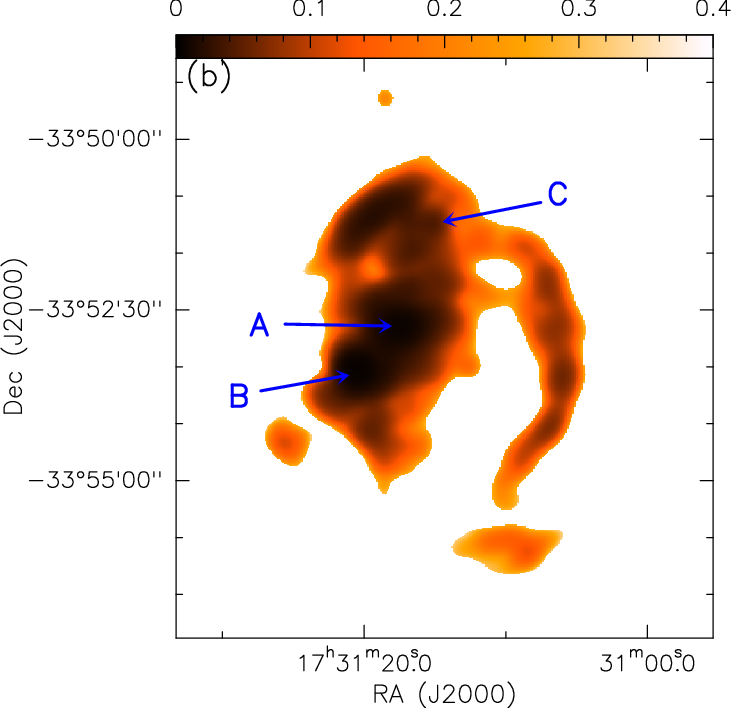}
\caption{{\it Left:} Spectral index map of the region associated with CS51.  
The low-resolution ($20\arcsec \times 20\arcsec$) 610-MHz contours are overlaid.
The contour levels are 0.5, 1, 2.5, 5, 10, 15, 25, 35, 45 and 80\% of the peak value (0.67~mJy/beam). The components A, B, and C are
marked. {\it Right:} The corresponding error map retrieved from the AIPS task {\tt COMB}.}
\label{radio_spdx}%
\end{figure*}

To probe the nature of the radio emission in the region associated
with the bubble, we construct a spectral index map from the 610 and 1300~MHz
data. This requires a pixel-by-pixel estimation of the spectral index, $\rm \alpha$, 
defined as $\rm S_{\nu} \propto \nu^{\alpha}$, where $\rm S_\nu$ is the flux density at the frequency $\nu$. Since we are interested in studying the variation of the
spectral index over the entire region associated with CS51, we need to sample the 
large-scale diffuse ionized emission. GMRT is not a scaled array between the 
observed frequencies implying that each frequency is sensitive to different spatial scales. Hence, to ensure that the contribution of diffuse 
emission is same at both frequencies, we generate a new map at 1300~MHz
from visibilities in the {\it uv} range of 0.1 -- 43~~K$\lambda$ which is consistent with the {\it uv} coverage in the 610-MHz map. The new 1300-MHz map is
regridded to the pixel size of 1\arcsec of the 610-MHz map. Both maps are
then convolved to a common lower resolution of $20\arcsec \times 20\arcsec$ to exclude any
small scale statistical fluctuation in the estimated spectral index values.
The above two steps are performed in the {\tt AIPS} environment using the tasks
{\tt LGEOM} and {\tt CONVL}. 
Retaining pixels above $5\sigma$ ($\rm \sigma \sim 1~mJy/beam$) in both maps, the spectral index map is obtained using the task {\tt COMB}. The error map is also retrieved from the task output to get the pixel uncertainties. The spectral index map and the corresponding error map is shown in Figure \ref{radio_spdx}. The
low-resolution 610-MHz contours are overlaid on the spectral index map.  

The spectral index values in the map vary between -1.75 to 1.5 and the
estimated errors are less than $\sim 0.4$. However, a careful examination
of the maps suggest that in the bubble interior the range is mostly
between -1.0 to 1.0 with errors less than $\sim 0.1$. There are few patches
showing extreme negative spectral index values and those are seen towards the eastern side of the component B and
also towards the circular void seen north-west of component A. The observed
range indicates the simultaneous presence of thermal free-free emission and non-thermal synchrotron emission which further suggest varying physical conditions. As discussed in \citet{1989ApJ...346L..85R}, when only 
free-free emission and absorption mechanisms are involved, the spectral index
lies between -0.1 and +2. Spectral indices $< -0.25$ are attributed to non-thermal emission \citep{{2016ApJS..227...25R},{1993ApJ...415..191C}}.
Radio spectral index maps of several Galactic \hii\ regions have shown similar co-existing thermal and non-thermal
components \citep{{2016A&A...587A.135R},{2016MNRAS.456.2425V},{2016AJ....152..146N}}. 
This has been attributed to shocks
induced by outflows and/or winds. Such colliding wind
interpretation is also cited for negative spectral indices seen towards Wolf-Rayet
nebula \citep{2005A&A...440..743B}.

In Table \ref{peak_specindx}, we list the spectral indices obtained from the map at the peak positions of A, B, and C. The table also gives the spectral index values obtained from the integrated flux densities. The later values for components B and C are fairly consistent with that obtained by \citet{2003A&A...407..957M}. 
From the above estimates we infer thermal free-free emission to be associated with components A and C whereas the emission from component B shows spectral signature for non-thermal emission. Compact radio sources exhibiting such large negative
spectral index have been discussed in \citet{2016A&A...587A.135R} and \citet{2014RMxAA..50....3R} where the latter study puts forth a viable scenario of 
colliding winds of a massive binary system which produces a shocked region 
between the components where electrons can reach relativistic speeds by
Fermi acceleration thus producing synchrotron emission. Similar interpretation
has been discussed by \citet{2011A&A...532A..92L} for the extended
non-thermal radio emission detected towards the W43 cluster. 
Protostars with an active magnetosphere could also give rise to such negative
spectral indices \citep{2013A&A...552A..51D}. Detailed study of component B
is required to ascertain its exact nature. 
 
\begin{table*}
\caption{Spectral indices of components A, B, and C.  } 
\label{peak_specindx}
\begin{tabular}{cccc}
\hline 
Components & \multicolumn{3}{c}{Spectral Index} \\
\cline{2-4} \\
& \multicolumn{2}{c}{GMRT 610 -- 1300~MHz} & ATCA 4.8 -- 8.6~GHz \\
\cline{2-4}
&  From map$^a$ & From Int. flux$^b$ & From Int. flux $^c$ \\
\hline
A & -0.01$\pm$0.006 & 0.19$\pm$0.06 & --\\
B & -0.33$\pm$0.004 & -0.78$\pm$0.06 & -0.87 \\
C & 0.19$\pm$0.040 & -0.14$\pm$0.07 & -0.16 \\
\hline
\end{tabular}
\\$^a$ relates to peak positions from spectral index map; $^b$ obtained from the integrated flux densities of the final convolved ($20\arcsec\times20\arcsec$) maps; $^c$ from \citet{2003A&A...407..957M}.
\end{table*}

\subsection{Identifying the ionizing stars}\label{ionizing}
A careful literature survey shows that there has been no identification of the
ionizing source responsible for the \hiir\ associated with the bubble CS51.
In this section we attempt to identify and propose the
possible exciting source(s) responsible for CS51.
As discussed in \citet{{2006ApJ...649..759C},{2007ApJ...670..428C}}, the association of
\hiirs\ with IR bubbles indicate that these are formed by O- and/or B-type 
of stars.  
The Lyman continuum photon flux estimated from the GMRT radio maps suggest a spectral type of O6V - O5.5.

Identification of candidate exciting star(s) has been carried out in several
bubbles based on MIR and NIR photometry of sources enclosed within the bubbles
\citep{{2009A&A...494..987P},{2010A&A...513A..44P},{2012A&A...544A..39J},{2013MNRAS.429.1386D},{2016ApJ...833...85B}}. We have followed the method adopted 
in a recent paper by \citet{2016MNRAS.458.3684O}. As we expect the candidate ionizing star(s) to be located
within the CS51 PDR, we consider 2MASS sources with good quality photometry (refer to
Section \ref{2mass_archive}) located within a radius of 2.1\arcmin~of the central
coordinate of CS51. Inspite of the proper motion of the exciting star(s), they are still likely to be within the PDR given that these IR bubbles are young objects. However, they
need not always be located at the geometrical centre of the bubble.

\begin{figure}
\centering
\includegraphics[scale=0.4]{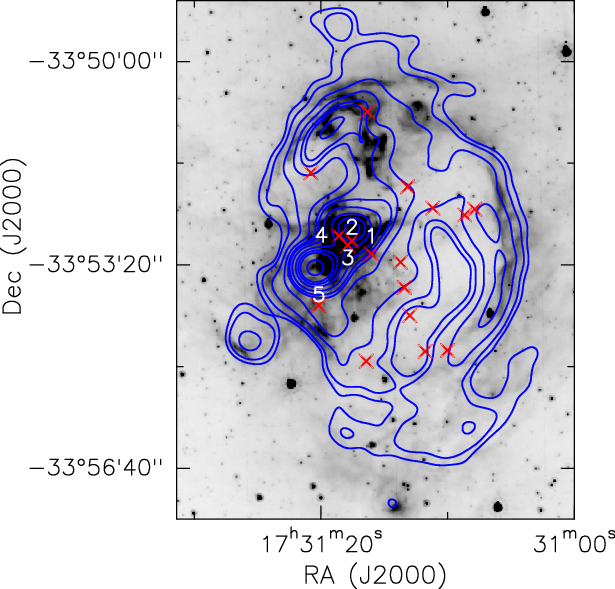}
\caption{This figure displays the 8-$\rm \mu m$ image on which is shown the distribution of O- and B-type stars (red crosses) 
in the bubble interior which satisfy the criteria discussed in \citet{2012A&A...543A.101C}. Low-resolution 610-MHz contours are overlaid. The contour levels are same as
in Figure \ref{radio_spdx}. The candidate ionizing stars are labeled as \#1, 2, 3, 4, and 5. Refer to the text for further discussion.}
\label{exciting_stars}%
\end{figure}

In our search for O stars within the CS51 PDR, we follow the BJHK colour criteria
of \citet{2012A&A...543A.101C}. The criteria adopted by these authors is highly
efficient in filtering out late-type contaminants. The B magnitudes of the retrieved sources are taken from the NOMAD1 catalog \citep{2004AAS...205.4815Z}. Since the bubble
CN20 studied by \citet{2016MNRAS.458.3684O} is at a similar distance as CS51, we
apply the same $\rm K_s$ magnitude cut of 11 to exclude out reddened early-type stars in 
the background of CS51. This is based on the assumption of $\rm A_v \sim 1$ per kpc of
foreground extinction. We identify 17 sources satisfying the criteria of \citet{2012A&A...543A.101C}. These  sources are marked in Figure \ref{exciting_stars} which shows the 8-$\rm \mu m$ image overlaid with 
the contours showing the 610-MHz radio emission. Except one, the sources fall in the region of the Class III population in the 2MASS colour-colour plot. The ones which have IRAC band magnitudes
are also seen to populate the region occupied by Class III objects \citep{2010A&A...513A..44P}. In their study of similar IR bubbles, \citet{{2016MNRAS.458.3684O},{2016ApJ...818...95L},{2016ApJ...833...85B}} have discussed the likely association
of the location of the ionizing star(s) and bright 24-$\rm \mu m$ emission in the bubble interior which also correlates with peak of the free-free radio emission. Following this
argument, we identify five candidates labeled 1, 2, 3, 4, and 5 in Figure \ref{exciting_stars}. Considering their associated visual extinctions from the 
2MASS $\rm (H - K_s)$ and $\rm (J - H)$ colour-colour plot, we derive their absolute K magnitudes and estimate the spectral type from \citet{2006A&A...457..637M}.
The details of these five sources are given in Table \ref{exciting_table}.
The location of star \#5 is towards the bubble periphery and
hence it is less likely to qualify as a candidate exciting source. Similarly, 
star \#2 falls in the Class II YSO region in the 2MASS colour-colour plot.
Thus, based on their nature and position inside the bubble, we identify sources \#1, 3 and 4 as the most promising candidates for ionization of CS51. Optical and
NIR spectroscopy of these candidate exciting sources are required to qualify
the above discussion.  

\begin{table*}
\centering
\caption{\small Exciting star candidates associated with CS51.}
\label{exciting_table}
\begin{tabular}{cccccccccccc}
\\ \hline 
\# & RA (J2000) & DEC (J2000) & B & J & H & K & $\rm A_V$ & $\rm M_J$ & $\rm M_H$ & $\rm M_K$ & Spectral type \\
 &$(^h~~^m~~~^s)$ & (~$^\circ~~\arcmin~~~\arcsec$)&(mag)&(mag)&(mag)&(mag)& &(mag)&(mag)&(mag)& \\
\hline 
1 & 17 31 16.01 & -33 53 09.27 & 17.51 & 10.19 & 9.09 & 8.72 & 10.19 & -6.30 & -6.31 & -6.04 & $>$ O3V \\
2 & 17 31 17.56 & -33 52 56.50 & 19.52 & 11.73 & 10.51 & 9.79 & 12.39 & -5.38 & -5.27 & -5.21 & $>$ O3V \\
3 & 17 31 17.88 & -33 53 00.72 & 16.07 & 12.28 & 10.96 & 10.28 & 12.92 & -4.98 & -4.92 & -4.78 & O4V -- O3V \\
4 & 17 31 18.58 & -33 52 51.46 & 17.91 & 11.91 & 10.93 & 10.60 & 9.22 & -4.30 & -4.30 & -4.04 & O6.5V -- O6V \\
5 & 17 31 20.17 & -33 53 59.88 & 18.61 & 12.13 & 11.13 & 10.75 & 9.53 & -4.17 & -4.15 & -3.93 & O7V -- O6.5V \\
\hline 
\end{tabular}
\end{table*}

\subsection{Emission from cold dust component} \label{herschel}
We study the nature of the cold dust emission using the 160 -- 500~$\rm \mu m$ 
{\it Herschel} data.
Line-of-sight average molecular hydrogen column density and dust temperature maps are generated by  pixel-wise modified blackbody fits. In generating these maps we have excluded the 70-$\rm \mu m$ emission following the discussion given in several studies \citep{{2010A&A...518L..98P},{2011A&A...535A.128B},{2010A&A...518L..98P},{2012A&A...542A..10A}}. 
These studies reason that the emission at 70~$\rm \mu m$ may not be optically thin and that there may be significant contribution from the warm dust components (e.g., very small dust grains, protostars) and hence cannot be modeled with a single
temperature gray body. Thus we have only four points mostly on the Raleigh-Jeans part to constrain the model. 

The initial steps involve converting the SPIRE map units from MJy sr$^{-1}$ to 
Jy pixel$^{-1}$ which is the unit of the PACS images. Subsequent to this, the 
160, 250, 350-$\rm \mu m$ images are convolved and regridded to the lowest resolution (35.7$\arcsec$) and largest pixel size (14$\arcsec$) of the 
500-$\rm \mu m$ image. The convolution kernels are taken from \citet{2011PASP..123.1218A}. The above steps are carried out using the {\it Herschel}
data reduction software HIPE\footnote{The software package for Herschel Interactive Processing Environment (HIPE) is the application that allows users to work with the Herschel data, including finding the data products, interactive analysis, plotting of data, and data manipulation.}.

We estimate the background flux, $\rm I_{bg}$, in each band from a relatively smooth and dark region devoid of bright, diffuse emission and filamentary structures.
This region is located at an angular distance of $\sim 20\arcmin$ from the bubble and centered at $\rm \alpha_{2000}=17^h32^m14.46^s$, $\delta_{2000}=-34^\circ01\arcmin21.90\arcsec$. 
The background value was estimated by fitting a Gaussian function to the 
distribution of individual pixels in the specified region. The fitting was carried 
out iteratively by rejecting the pixels having values outside $\pm$2$\sigma$ till the fit converged \citep{{2011A&A...535A.128B},{2013A&A...551A..98L}}. We have used the same region for the determination of background offset in all the bands. $\rm I_{bg}$ is
estimated to be -0.82, 2.81, 1.27, and 0.45 Jy pixel$^{-1}$ at 160, 250, 350, and 500~$\rm \mu m$, respectively. The negative flux value at
160~$\rm \mu m$ is due to the arbitrary scaling of the PACS images. 

\begin{figure*}
\centering
\includegraphics[scale=0.25]{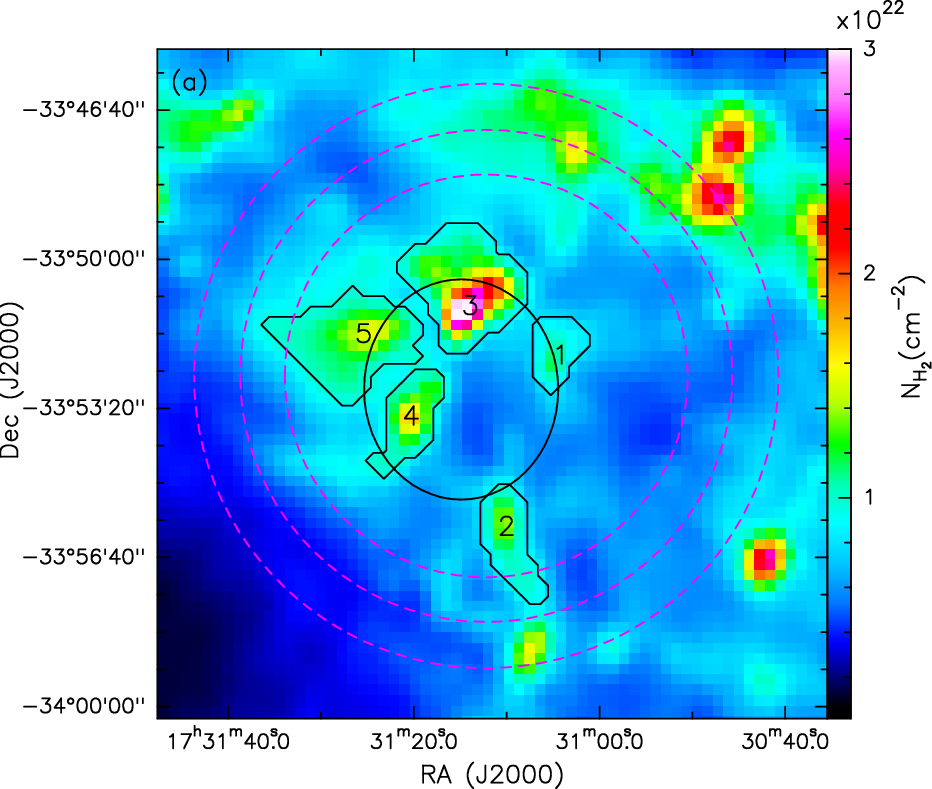} 
\includegraphics[scale=0.25]{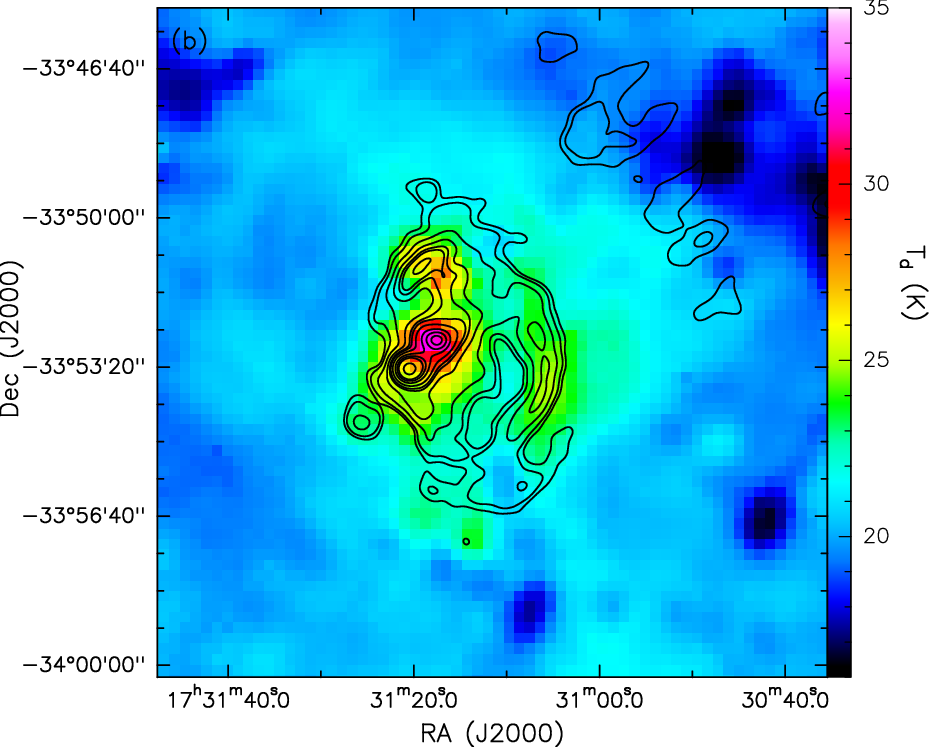}
\includegraphics[scale=0.25]{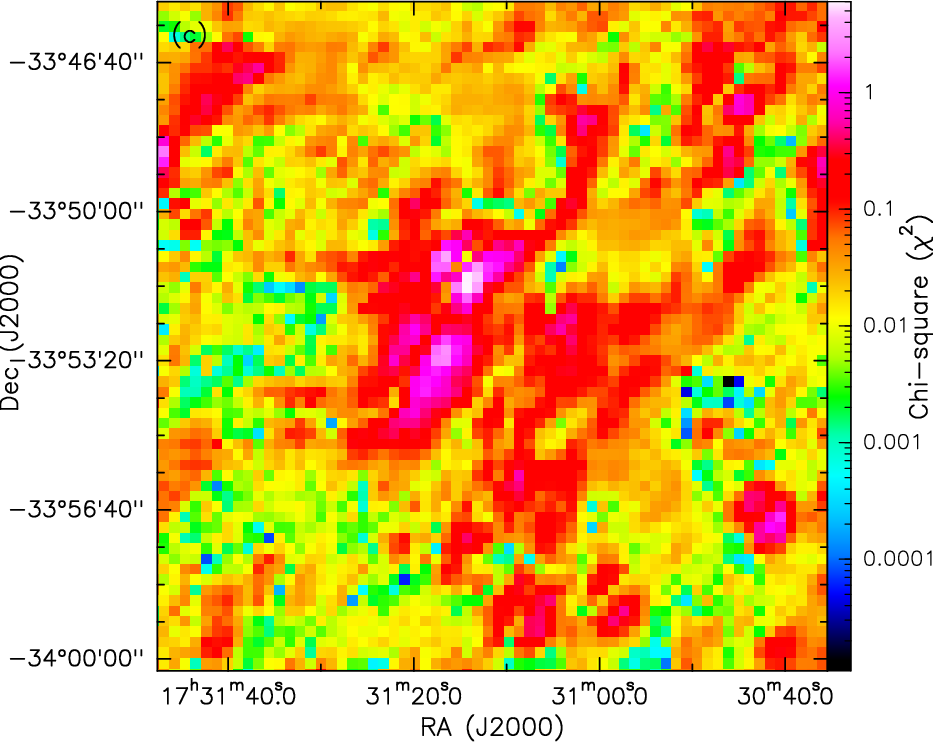}
\caption{Column density (a) and dust temperature (b) and chi-square (c) maps generated using {\it Herschel} FIR data are shown. In plot (a), the black ellipse
shows the extent of the bubble and the five identified cold dust clumps are shown as black contours. This plot also shows three concentric circles (dashed magenta) which are used in the study of column density PDFs (see Section \ref{prob_den}).
610-MHz radio contours are overlaid in plot (b) to correlate the dust temperature with the ionized gas distribution. The contour levels are same as in Figure \ref{radio_spdx}. }
\label{dust_temp}%
\end{figure*}

We model the FIR emission with a modified blackbody that takes  into  consideration  
the optical depth, and the dust emissivity. As mentioned earlier, this is done pixel-wise adopting the 
following functional form \citep{{2011A&A...535A.128B},{2012MNRAS.426..402F},{2013A&A...551A..98L},{2015MNRAS.447.2307M}}. 
\begin{equation}
\rm S_{\nu}(\nu) - I_{bg}(\nu) = B_{\nu}(\nu,T_{d})\ \Omega\ (1-e^{-\tau_{\nu}})
\end{equation}
and 
\begin{equation}
\rm \tau_{\nu} = \mu_{\rm H_{2}}\ m_{\rm H}\ \kappa_{\nu}\ N({\rm H_{2}}) 
\end{equation}
where, $\rm S_{\nu}(\nu) $ is the observed flux density, $\rm I_{bg}(\nu) $ is the 
background flux, $\rm B_{\nu}(\nu,T_{d}) $ is the Planck's function, $\rm T _{d} $ 
is the dust temperature, $\rm \Omega $ is the solid angle (in steradians) from where 
the flux is obtained (solid angle subtended by a 14$\arcsec \times 14\arcsec~$pixel), $\rm 
\mu_{\rm H_{2}}$ is the mean molecular weight, $\rm m_{\rm H}$ is the mass of hydrogen atom,
$\rm \kappa_{\nu}$ is the dust opacity and $\rm N$(H$_{2} $) is the column density. 
We have assumed a value of 2.8 for  $\mu_{\rm H_{2}}$ \citep{2008A&A...487..993K}. 
The dust opacity $\rm \kappa_{\nu} $ 
is defined to be $\rm \kappa_{\nu} = 0.1~(\nu/1000~{\rm GHz})^{\beta}~{\rm cm^{2}/g} 
$, where, $\beta$ is the dust emissivity spectral index \citep{{1983QJRAS..24..267H}, {1990AJ.....99..924B}, {2010A&A...518L.102A}}. The SED fitting was carried 
out using non-linear least square Levenberg-Marquardt algorithm, in which  $\rm 
T_{d}$ and $\rm N$(H$_{2})$ are kept as free parameters. 
Given the limited number of data points, we prefer to fix the value of $\beta$ to 2
\citep{{1983QJRAS..24..267H}, {1990AJ.....99..924B}, {2010A&A...518L.102A}} which is also a typical value used in studies related to IR dust bubbles \citep{{2012A&A...542A..10A},{2016ApJ...818...95L},{2016MNRAS.458.3684O}}.
We have used a 
conservative 15\% uncertainty on the background subtracted flux densities 
\citep{2013A&A...551A..98L}. The generated column density and temperature 
maps are shown in Figure \ref{dust_temp}, alongwith the corresponding $\chi^2$ map. We have overlaid the low resolution ($\rm 20\arcsec \times 20\arcsec $) 610~MHz radio map on the dust temperature map to correlate the ionized and dust components. 

As seen in Figure \ref{dust_temp}, the pixel fits are good with $\chi^2 < 6$ in
the map.  
The column density map shows a fragmented shell-like structure with low values tracing the inner region of the bubble. The bubble rim is seen to contain clumpy regions of enhanced column density. One clump is located relatively close to the centre of
the bubble coincident with the radio peaks. Apart from this clump, the rest
fit well into a picture of fragmented swept-up molecular shell. The peak column density (3.6$\times 10^{22}\rm cm^{-2}$) is located towards the northern part of the bubble. Several clumpy structures and high density 
regions are also seen on the periphery and beyond the bubble. Correlating the
ionized emission with the dust temperature map, we see high dust temperature
towards the three components and the entire region associated with the
bubble CS51 is clearly warmer than the surrounding region. 

As shown in Figure \ref{dust_temp}, five cold dust clumps associated
with CS51 are detected from the column density map. Instead of using the
visual inspection followed by fitting ellipses to characterize the clumps \citep{2016ApJ...818...95L}, we use the {\it astrodendro} algorithm \citep{{2008ApJ...679.1338R},{2009Natur.457...63G}}. In doing so, we set a 
threshold column density of 8.0$\rm \times 10^{21}\ cm^{-2}$ and detect 
clumps above this by specifying a minimum of 5 pixels for a potential
clump.  

Mass of the dust clumps is calculated using the following equation
\begin{equation}
\rm M_{\rm clump} = \mu_{\rm H_{2}} m_{\rm H} A_{\rm pixel} \Sigma N ({\rm H_{2}}) 
\label{clump_mass1}
\end{equation}
where, $\rm m_{\rm H} $ is the mass of hydrogen, $\rm A_{\rm pixel} $ is the pixel area in $\rm cm ^{2} $, $ \mu_{\rm H_{2}} $ is the mean molecular weight and $\rm \Sigma N ({\rm H_{2}})$ is the integrated column density. 
The clump apertures retrieved from {\it astrodendro} are used to
estimate the physical sizes ($\rm r=(A/\pi)^{0.5}$; \citep{2010ApJ...723L...7K}). The derived physical properties of the dust clumps
are listed in Table \ref{clump_properties}. The volume number density is derived as, $\rm n_{H_2}= 3\Sigma \,NH_2\,/4r$, r being the radius of the clump.
The two clumps detected from the 1.2-mm emission by \citet{2006A&A...447..221B}
are likely to be associated with our Clump 3 and Clump 4. However, the masses
derived in this paper are larger by a factor of 2 -- 7 compared to the estimates
of \citet{2006A&A...447..221B}. The possible reasons for this could be the different $\kappa_{\nu}$ and dust temperature
values used by these authors and/or the
large sizes of the clumps retrieved from the column density map. One also
cannot exclude the effect of flux loss associated with ground based observations \citep{2017A&A...602A..95L}.  

\begin{table*}
\centering
\caption{Physical parameters of the clumps associated with CS51. 
The peak position, radius, mean dust temperature, mean column density, mass and
volume density for each clump is listed.}
\label{clump_properties}
\begin{tabular}{ccccccccc}
\\ \hline
Clump No. & RA (2000) & DEC (2000) & Radius & Mean $\rm T_{d}$ & Mean $\rm N(\rm H_{2})$ & $\rm \sum N(\rm H_{2}$) & Mass & $\rm n_{H_2}$ \\
& $(^h~~^m~~~^s)$ & (~$^\circ~~\arcmin~~~\arcsec$) & (pc) & (K) & ($\rm \times 10^{21}cm^{-2}$) & ($\rm \times 10^{23}cm^{-2}$) & (M$_{\odot}$) & ($\rm \times 10^{3}cm^{-3}$) \\
\hline
1 & 17 31 05.27 & -33 52 25.35 & 1.0 & 22.7 & 9.4 & 2.3 & 810 & 2.7  \\
2 & 17 31 09.68 & -33 55 54.26 & 1.2 & 20.7 & 10.5 & 3.8 & 1340 & 2.5 \\
3 & 17 31 15.29 & -33 51 15.10 & 1.9 & 22.9 & 18.4 & 13.3 & 4600 & 2.2 \\
4 & 17 31 19.75 & -33 53 35.20 & 1.2 & 26.8 & 10.1 & 4.2 & 1460 & 2.9 \\
5 & 17 31 24.48 & -33 51 15.10 & 1.9 & 20.9 & 11.4 & 9.8 & 3420 & 1.7 \\ 
\hline                
\end{tabular}
\end{table*}

\subsection{Associated stellar population} \label{yso}
To understand the star formation activity associated with CS51, we study the 
nature of the stellar population. We use {\it Spitzer} GLIMPSE and MIPSGAL data to identify and classify potential YSOs towards the bubble. 
For this we choose a region of radius 5$\arcmin$ centered on the bubble.
We retrieve 1060 sources with photometric data in all the IRAC bands. We also
explore the catalog of \citet{2015AJ....149...64G} to include 24~$\rm \mu m$
point sources which have 3.6~$\rm \mu m$ photometry. 
From the distribution of IRAC sources extracted from the catalog, we see a dearth of
point sources towards regions of bright MIR emission located in the
bubble centre and periphery. 
Hence, to ensure a more complete sample of sources associated
with the bubble we also make use of good photometric data from the 2MASS Point Source Catalog. A total of 1418 sources have good quality data in all three
$\rm JHK_s$ bands. The identification and classification of YSOs are carried out using the following procedures.

\begin{figure*}
\centering
\includegraphics[scale=0.3]{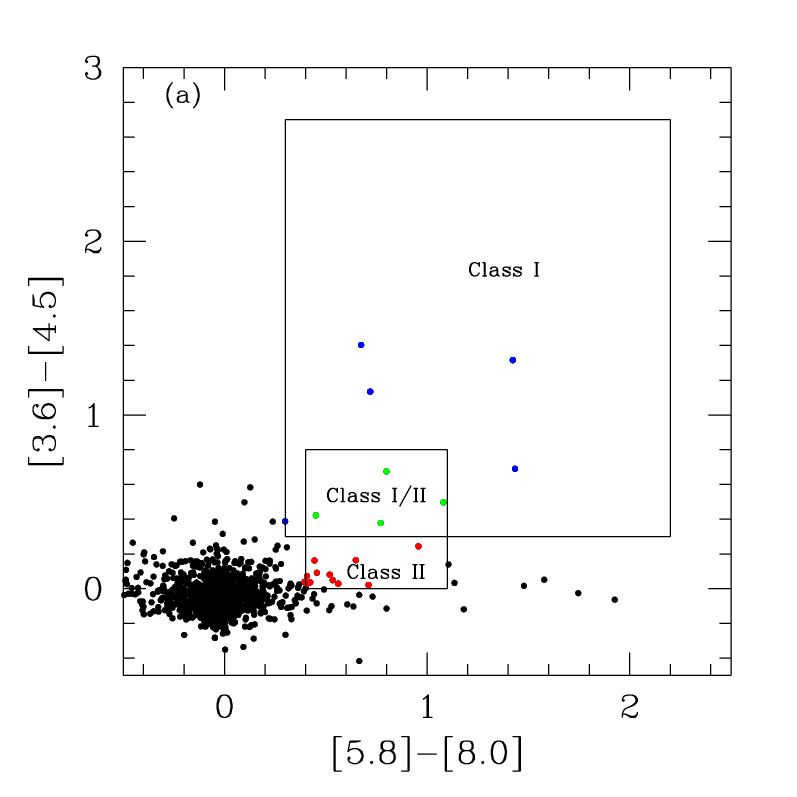}
\includegraphics[scale=0.3]{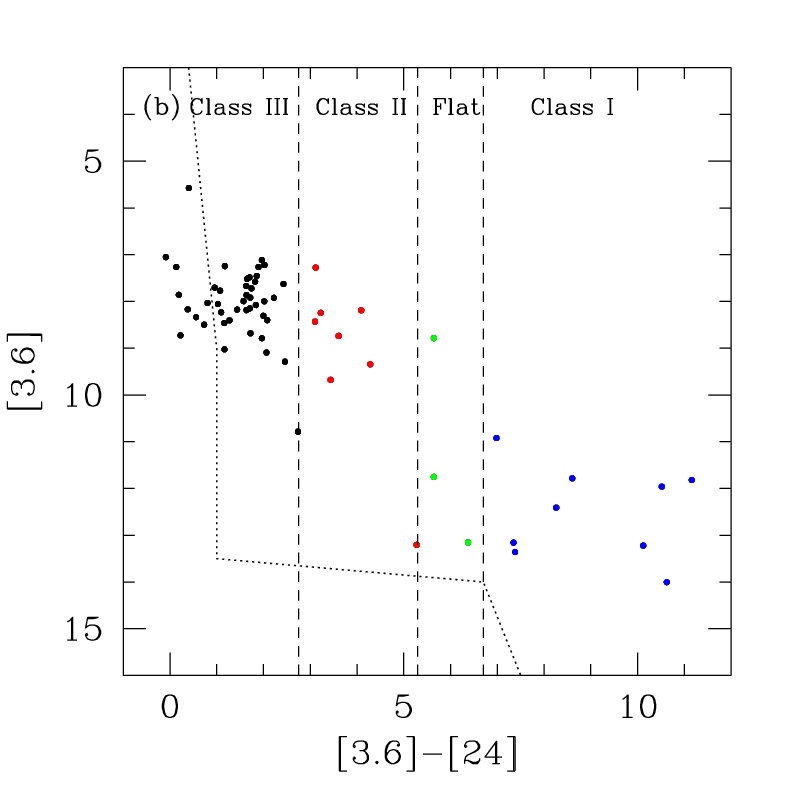}
\includegraphics[scale=0.3]{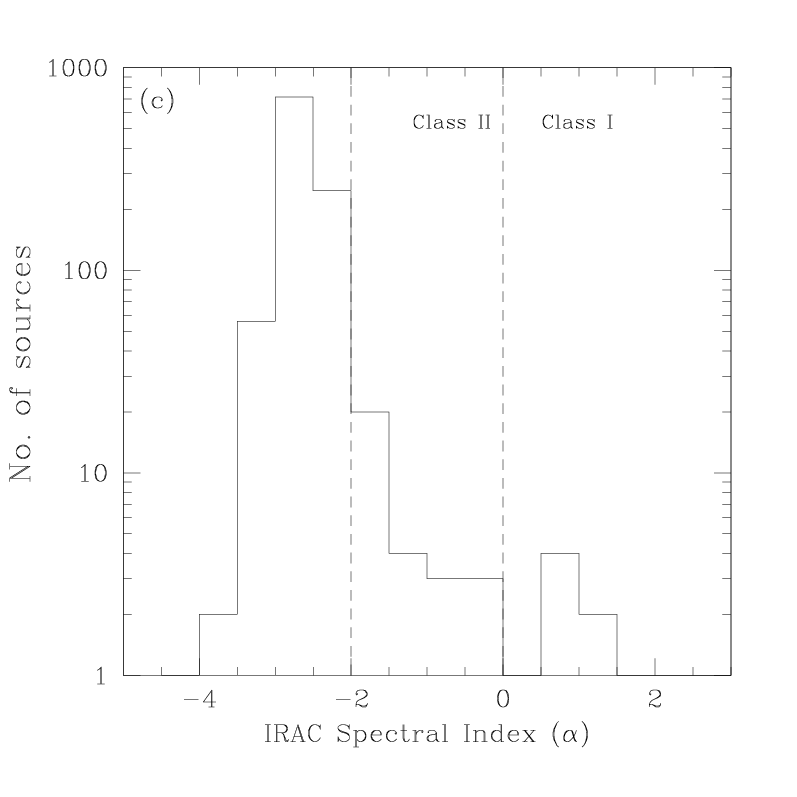}
\includegraphics[scale=0.3]{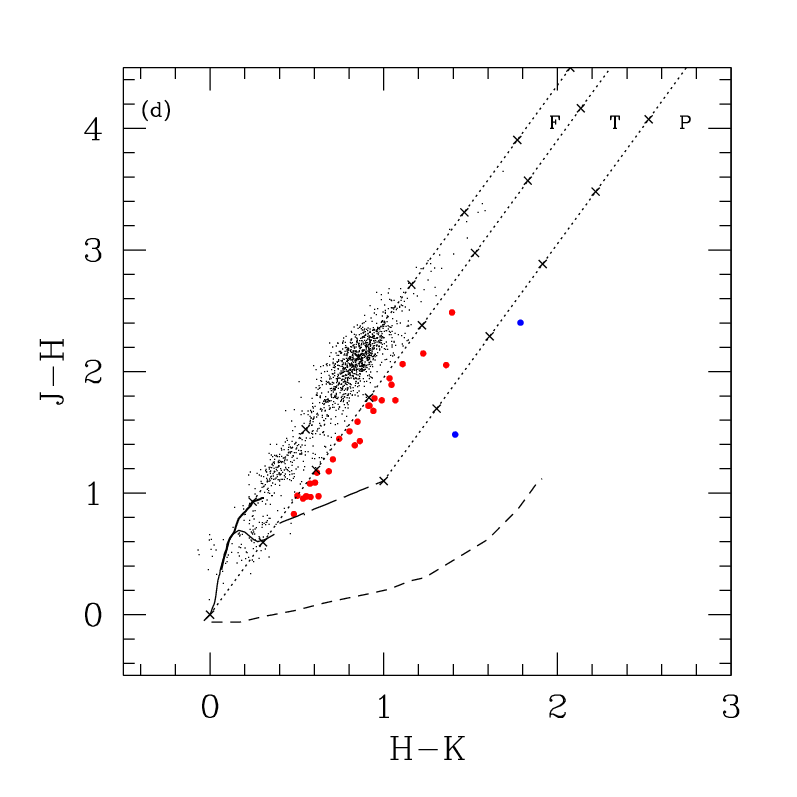}
\caption{(a) IRAC CCP plot following the method of \citet{2004ApJS..154..363A}. The boxes on the plot, adapted from \citet{2007A&A...463..175V} shows the region for YSOs of different evolutionary classes. (b) [3.6] - [24] vs [3.6] CMP on which the vertical lines adopted from \citet{2010ApJ...720...46G} and \citet{2011ApJS..196....4R} demarcate the regions for different classes of YSOs. (c) Histogram plot of the variation of number of sources with IRAC spectral index. The regions for Class I and Class II sources labeled on the plot are adapted from \citet{2008ApJ...682..445C}. (d) 2MASS J-H vs H-K CCP , in which the three separate regions are labeled as `F', `T', and `P' (refer text). The colour coding for YSOs are as follows: Class I -- blue, Class II -- red and Class I/II -- green. }
\label{cs51_yso}%
\end{figure*}

\begin{enumerate}

\item[(a)] This scheme uses the IRAC colors for the classification of YSOs as discussed in \citet{2004ApJS..154..363A}. Class I (sources dominated by protostellar envelope emission) and Class II (sources dominated by protoplanetary disk) are segregated based on their location on the [3.6] - [4.5] vs [5.8] - [8.0] colour-colour plot (CCP), shown in Figure \ref{cs51_yso}(a). The boxes on the plot demarcating the location of Class I and Class II sources, are adopted from \citet{2007A&A...463..175V}. Following this classification scheme, we have identified 21 YSOs, out of which 5 are Class I, 12 are Class II and 4 are Class I/II.    

\item[(b)] In this method, we used the [3.6] - [24] vs [3.6] colour-magnitude plot (CMP) and the criteria discussed in \citet{2010ApJ...720...46G} and \citet{2011ApJS..196....4R} for the classification of YSOs. Out of 105 detected 
 24~$\rm \mu m$ point sources, 63 have 3.6~$\rm \mu m$ counterparts in the GLIMPSE highly reliable catalog. Figure \ref{cs51_yso}(b) shows the colour-magnitude plot (CMP), where the vertical lines separate the YSOs of different evolutionary classes. The dotted curve on the plot marks the boundary of contaminated sources such as galaxies and disk-less stars. Following this method, 20 YSOs are identified, out of which 9 are Class I, 8 are Class II, and 3 are flat-spectrum YSOs.

\item[(c)] In this procedure, YSOs are classified based on their IRAC spectral index. The IRAC spectral index ($\alpha = d\ {\rm log}(\lambda F_{\lambda}) / d\ {\rm log}(\lambda)$; \citealt {1987IAUS..115....1L}) is estimated for each source by a linear regression fit. Then the YSOs are classified into Class I $(\alpha > 0)$ and Class II $(-2\leqslant \alpha \leqslant 0)$, following the classification scheme of \citet{2008ApJ...682..445C}. The distribution of YSOs following this classification scheme is shown in Figure \ref{cs51_yso}(c). Here, we identify 38 YSOs out of which 8 are Class I and 30 are Class II sources. 

\item[(d)] $\rm JHK_s$ CCPs have been efficiently used by various authors \citep{{2002ApJ...565L..25S},{2004ApJ...608..797O},{2004ApJ...616.1042O},{2006A&A...452..203T},{2012A&A...544A..39J},{2015MNRAS.447.2307M}} to
identify YSOs. The CCP for the NIR sources associated with bubble CS51 is shown in Figure \ref{cs51_yso}(d). The loci of main sequence (thin line) and giants (thick line) are taken from \citet{1988PASP..100.1134B}. The classical T Tauri locus shown as a long-dashed line is taken from \citet{1997AJ....114..288M}. The parallel dotted lines are the reddening vectors, on which the cross marks are the visual extinction placed at intervals of 5 mag. The short-dashed line represents the locus of Herbig AeBe sources is taken from \citet{1992ApJ...393..278L}. We have assumed the interstellar reddening law from \citet{1985ApJ...288..618R}. In this plot all the colours and curves are converted into \citet{1988PASP..100.1134B} system. For better classification this plot is divided into three regions. The sources in `F' region are field stars or Class III or Class II stars with small NIR excess. Sources in `T' region are mainly classical T-Tauri or Class II stars. Sources in the `P' region are mostly Class I YSOs. To classify YSOs, we have considered sources falling above T-Tauri locus, since there could be 
overlap of Class I and Class II sources with Herbig AeBe sources. Following this method we identify 31 YSOs, out of which 2 are Class I YSOs and 29 are Class II YSOs. Given the distance to CS51, the 2MASS photometry would be severely affected
foreground interstellar extinction thus increasing the contamination from foreground
field stars.

\end{enumerate}

\begin{figure}
\centering
\includegraphics[scale=0.3]{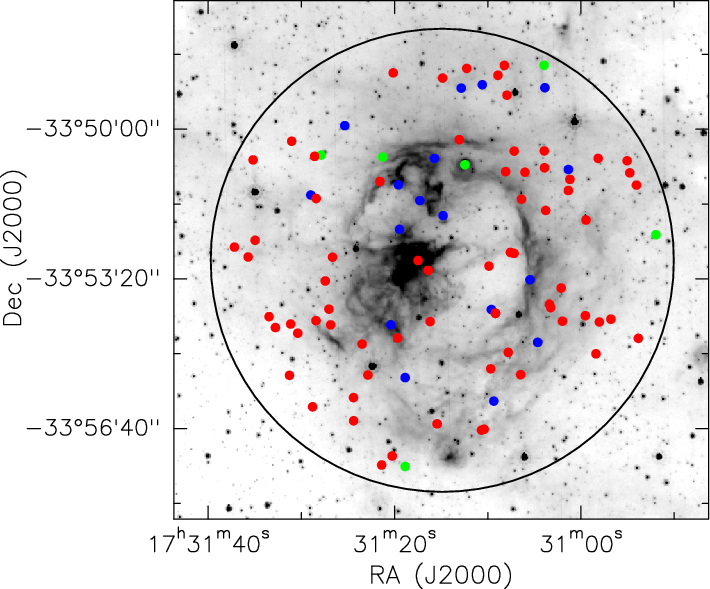}
\caption{Gray scale is 8.0-$\rm \mu m$ image on which the distribution of YSOs are shown. The colour coding for YSOs are as follows: Class I -- blue, Class II -- red and Class I/II -- green.}
\label{cs51_yso_dis}%
\end{figure}

Compiling the YSOs identified in the four methods, we have 17, 70, and 6 Class I,
Class II, and Class I/II sources, respectively in the region of 5\arcmin~radius
centered on CS51. For YSOs identified in more than one scheme, methods (a), (b), 
(c), and (d) are given preference in this order to assign their class.
Figure \ref{cs51_yso_dis} shows the distribution of the 
identified YSOs on the 8-$\rm \mu m$ image and Table \ref{yso_table} lists
the coordinates, NIR and MIR magnitudes and classification. The spatial distribution
of YSOs do not show any particular pattern except a dearth in the bubble interior and an overdensity of Class I YSOs towards the northern periphery of the bubble.
Two Class II YSOs are placed in projection towards the central clump thus hinting
at a more evolved region compared to the clump towards the northern rim. 
Without spectroscopic confirmation it is difficult to conclude on the physical 
association of the identified YSOs with CS51. 

\subsection{Molecular line emission towards CS51}
\label{mol_emm}
In order to investigate the properties of the clumps associated with the bubble, we have used molecular line data from the MALT90 survey. The available data covers two of the {\it Herschel} clumps (\# 3 and \# 4). Study of these molecular lines will
enable us to probe the physical, chemical properties and throw light on the evolutionary states of these dense star-forming clumps. 
Of the 16 molecular-line transitions covered in the MALT90 survey, only five molecular species ($\rm C_2H$, HCN, $\rm HCO^+$, HNC, and $\rm N_2H^+$) are detected
towards these two clumps. The details of the transitions taken from \citet{2011ApJS..197...25F} and \citet{2014A&A...562A...3M} are listed in
Table \ref{spectra_detail}. Both the papers, especially the latter, give
a nice review on the environment in which these molecules exist
and the physical conditions required for their formation. 

\begin{figure*}
\centering
\includegraphics[scale=0.2]{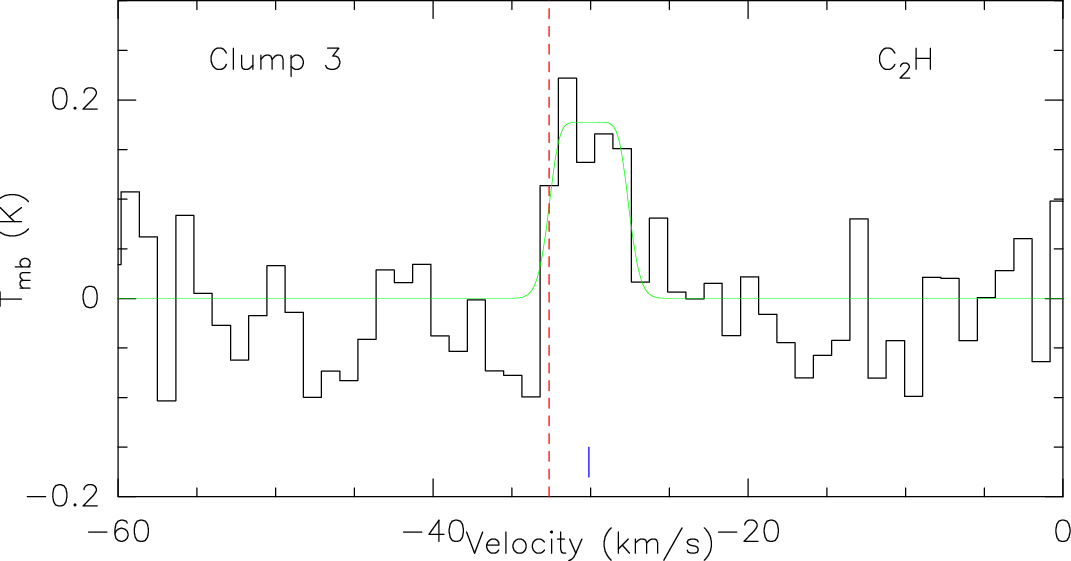}
\includegraphics[scale=0.2]{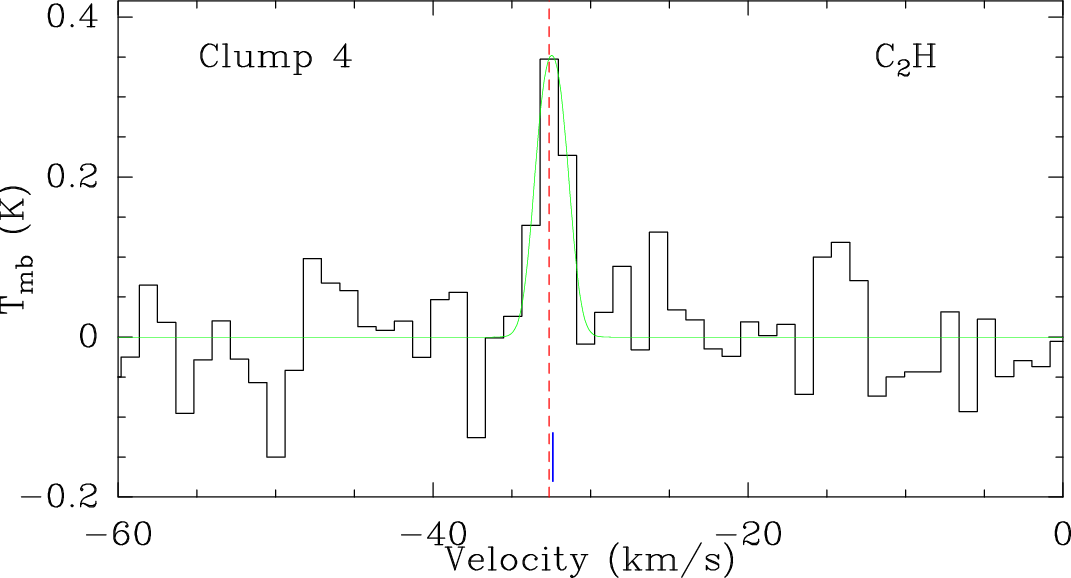}
\includegraphics[scale=0.2]{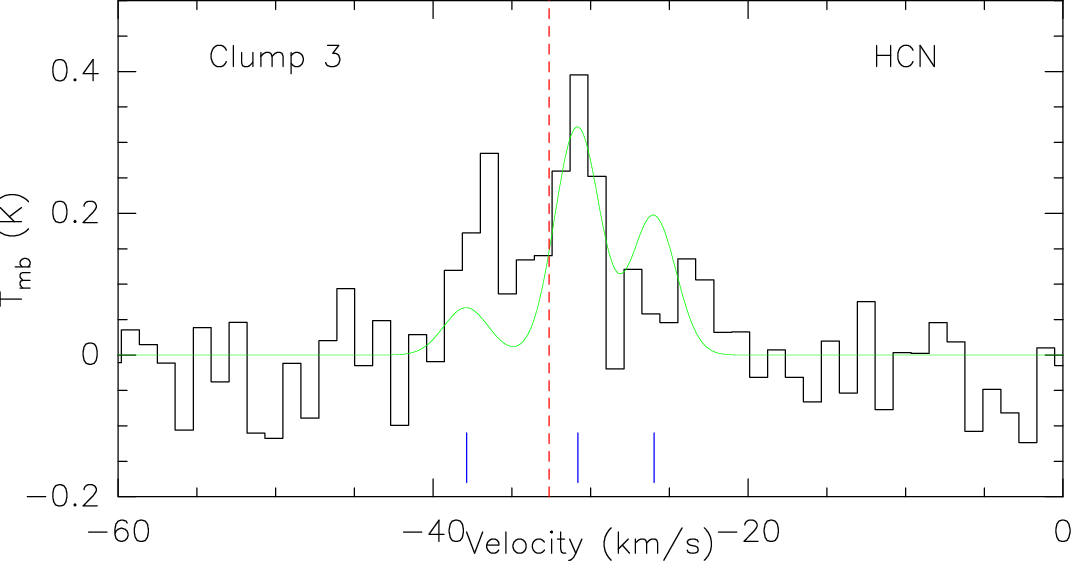}
\includegraphics[scale=0.2]{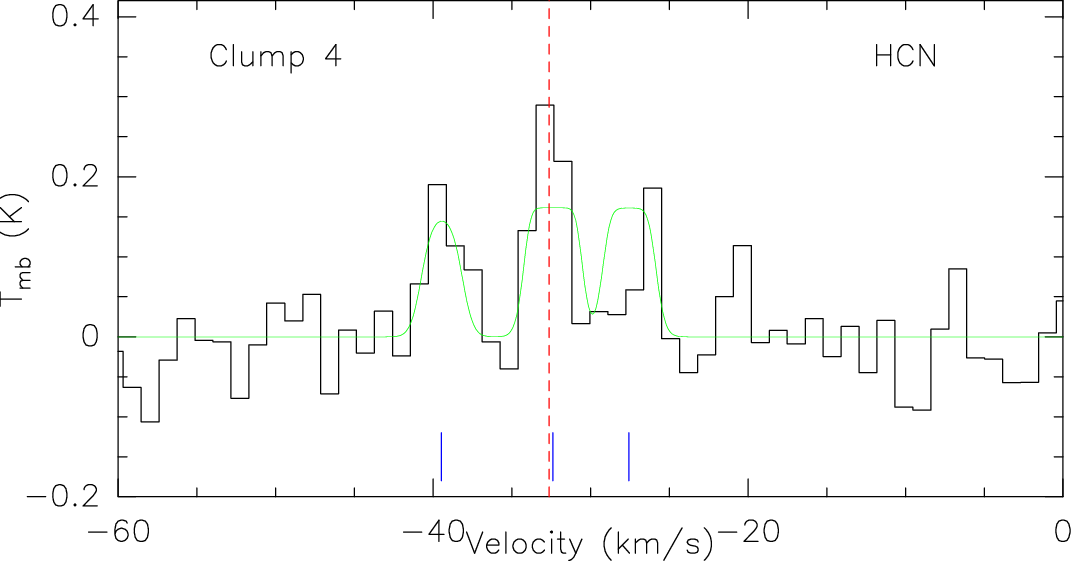}
\includegraphics[scale=0.2]{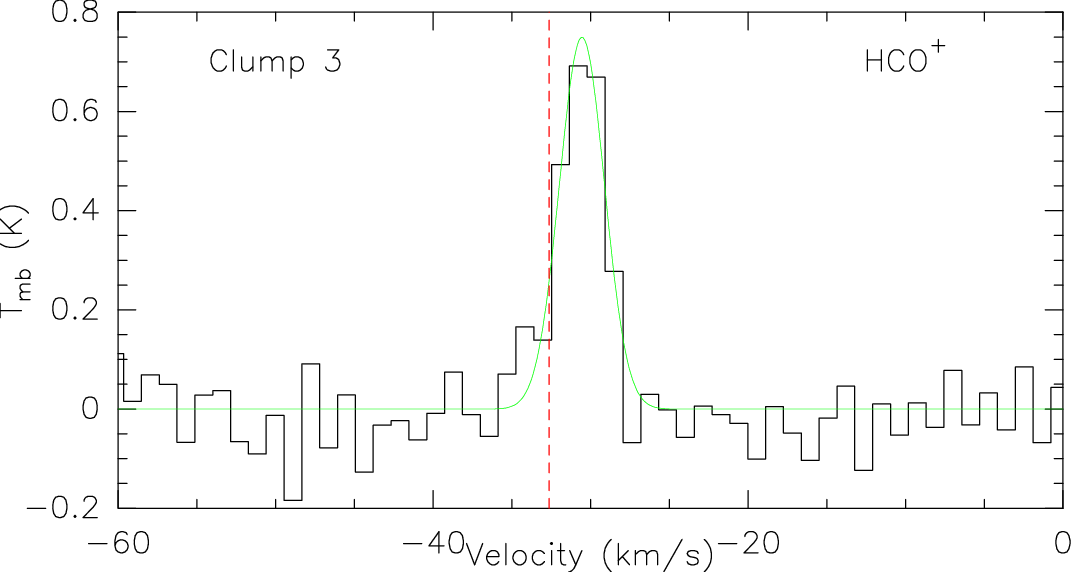}
\includegraphics[scale=0.2]{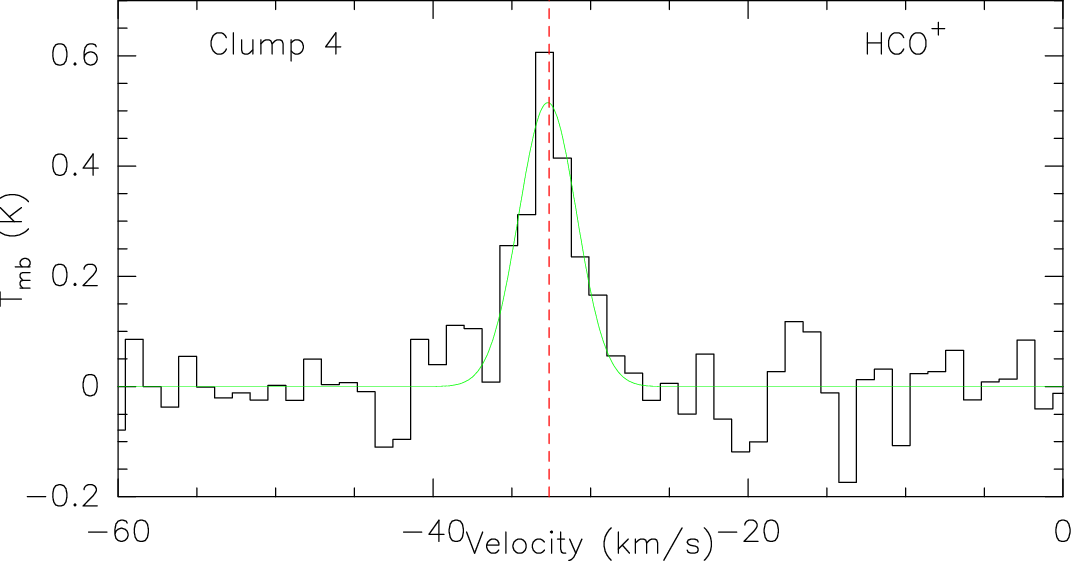}
\includegraphics[scale=0.2]{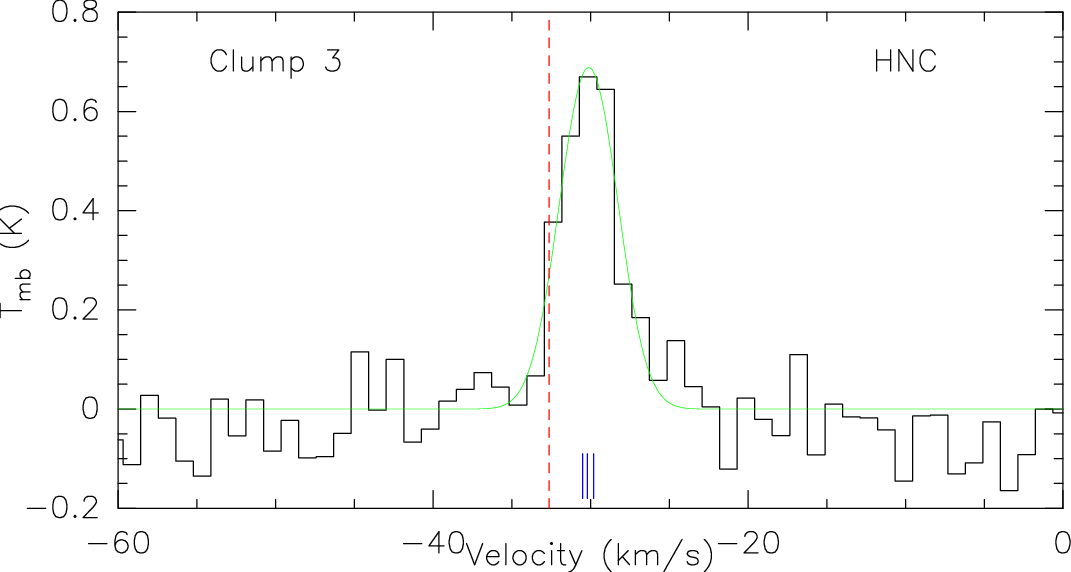}
\includegraphics[scale=0.2]{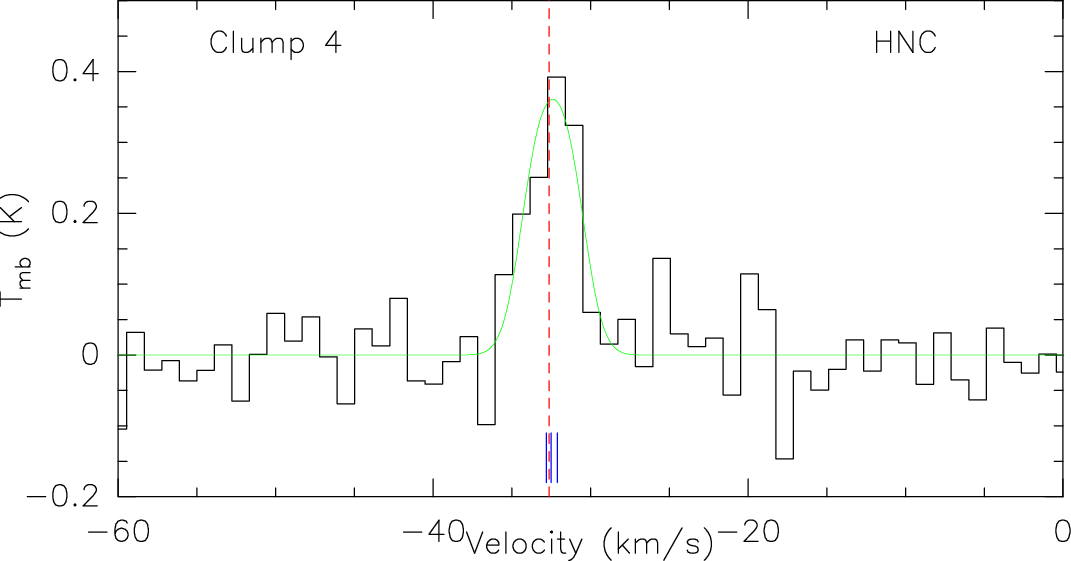}
\includegraphics[scale=0.2]{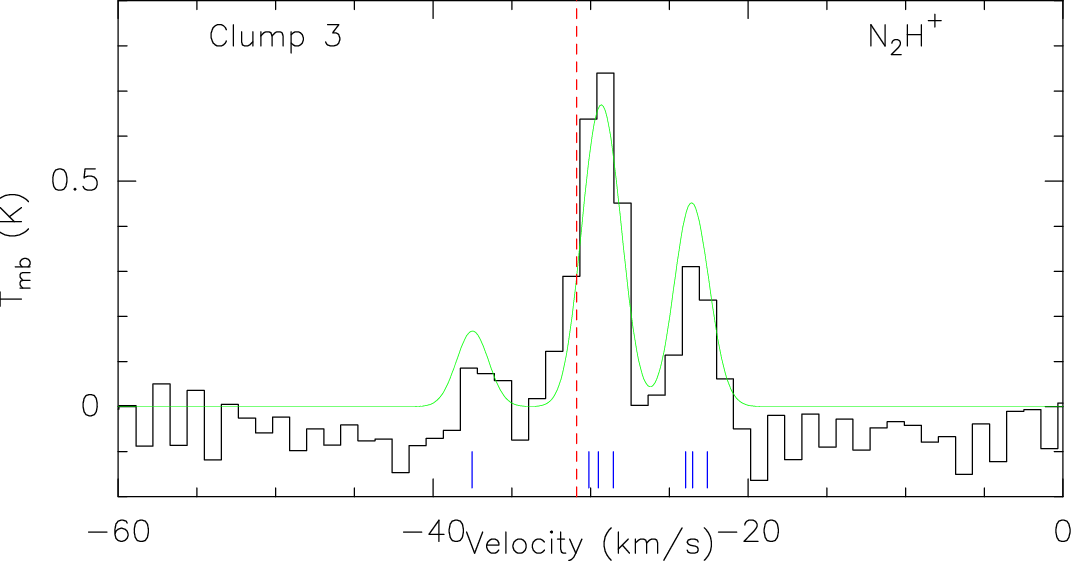}
\includegraphics[scale=0.2]{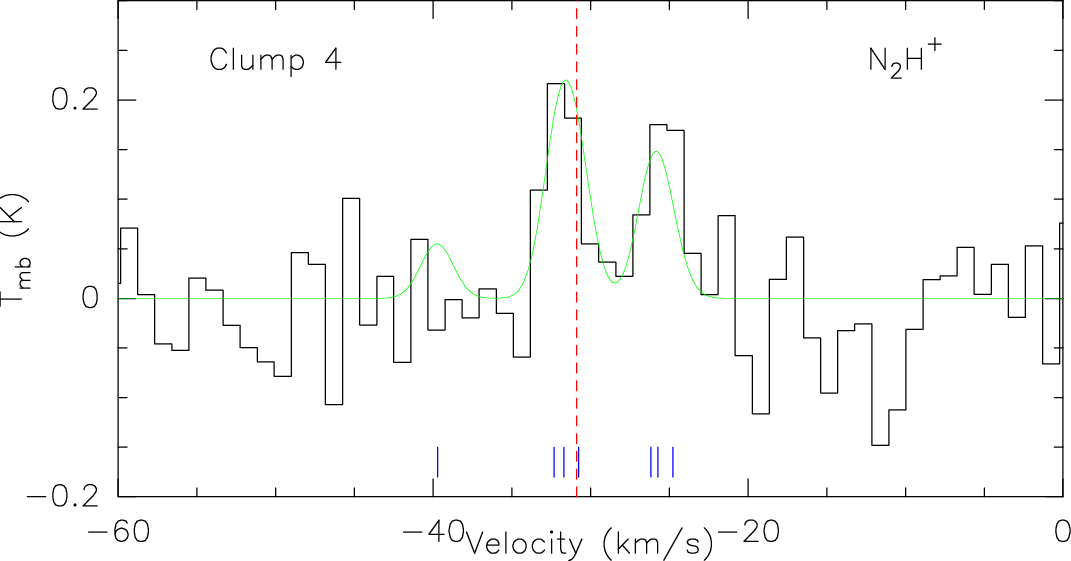}
\caption{Spectra toward Clumps 3 and Clump 4 associated with CS51. The green curves show the {\tt hfs} and Gaussian (for $\rm HCO^+$) fits to the spectra. The dashed, vertical red line marks the average peak velocity ($\rm V_{LSR}$) of all 
detected molecules of Clump 4 that is used as the systematic velocity. The 
hyperfine components of the respective molecules are indicated with blue lines.}
\label{spectra}%
\end{figure*}

\begin{figure*}
\centering
\includegraphics[scale=0.3]{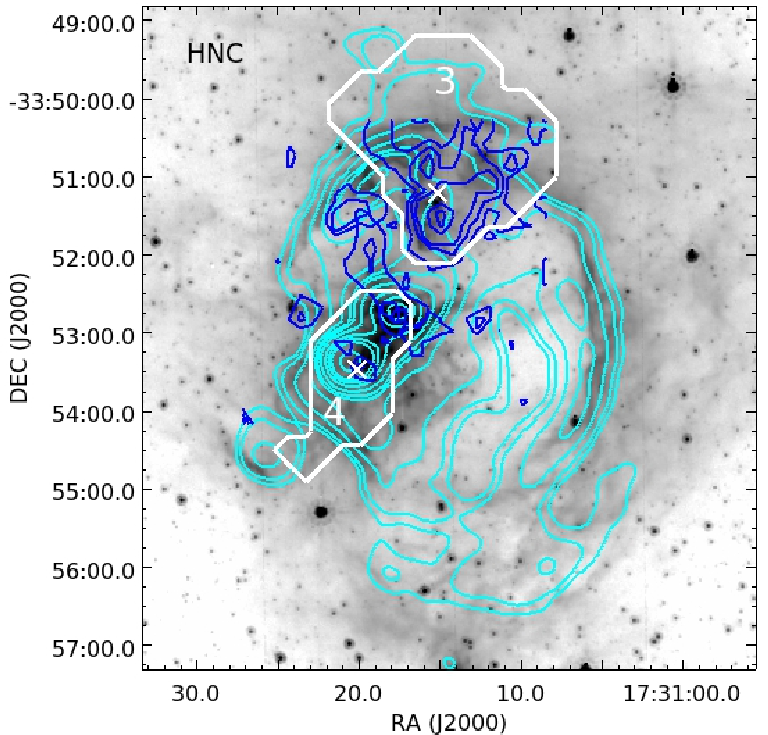}
\hspace{1cm}
\includegraphics[scale=0.3]{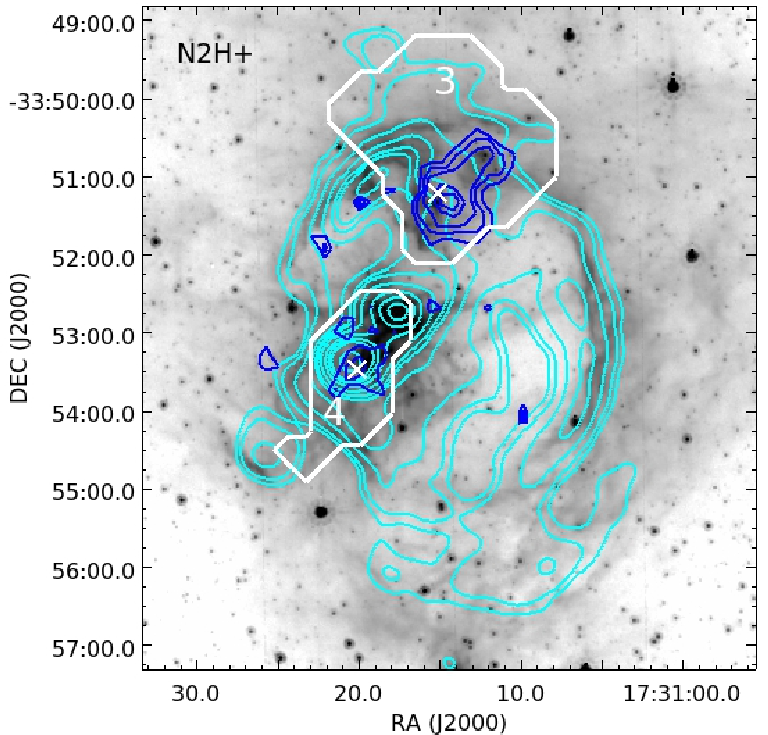}
\caption{{\it Spitzer} 8-$\rm \mu m$ is shown in gray scale. Blue contours are the integrated intensity maps. The contour levels starts from 3$\sigma$, where $\sigma$ = 0.4~$\rm K\ km\ s^{-1}$ for both the molecules.  610-MHz radio contours in cyan shows the distribution of ionized gas. The contour levels are same as in Figure \ref{radio_spdx}. The white crosses are the pointing of MALT90 observation. The retrieved clump apertures (see section \ref{herschel}) are also shown.}
\label{integrated}%
\end{figure*}

Figure \ref{spectra} shows the observed line-spectra at the peak positions for the two clumps. 
We use the hyperfine structure ({\tt hfs}) method of CLASS90 to fit the observed
spectra for
$\rm C_2H$, HCN, HNC, and $\rm N_2H^+$ transitions and retrieve the line parameters. As described in Table \ref{spectra_detail}, the molecule 
$\rm HCO^+$ has no hyperfine structure thus a single Gaussian profile
was used to fit the spectra. Figure \ref{spectra} also shows the {\tt hfs} and 
Gaussian (for $\rm HCO^+$) fits for the detected transitions. The positions of the hyperfine components are 
marked in the plots. The retrieved parameters are peak velocity ($\rm V_{LSR}$), width of velocity ($\Delta V$), main beam temperature ($\rm T_{mb}$), and velocity integrated intensity ($\rm \int T_{mb} dV$),  which are listed in Table \ref{spectra_param}. Beam correction 
is applied to the antenna temperature to obtain the main beam 
temperature, $\rm T_{mb} = T_{A}/ \eta_{mb}$ \citep{2014ApJ...786..140R} with
an assumed value of $\eta_{mb}$ = 0.49 \citep{2005PASA...22...62L}. As revealed in the 
figure, the signal-to-noise ratio of the observed spectra are less than
optimal with several hyperfine components being marginally detected and few others
fitting to appreciably large widths (see $\rm C_2H$ profile for Clump 3 and HCN profile for Clump 4). 
In view of this, it should be noted that the retrieved parameters would have significant uncertainties associated. 
The 0th moment (velocity integrated) contours of two
molecules, HNC and $\rm N_2H^+$, are shown in Figure \ref{integrated} overlaid on the {\it Spitzer} 8~$\rm \mu m$ image. The maps of the other molecules have poor signal-to-noise ratio and hence are not presented. Both molecules show extended emission towards Clump 3 as compared to Clump 4. The plots also show the
610-MHz contours that enable us to correlate the emission from molecular and
ionized gas.

\begin{table*}
\centering
\footnotesize
\caption{\small Summary of detected spectral-line transitions. $E_u$ and $n_{crit}$ are the excitation energy and critical density for the transitions. These are calculated adopting values from the Leiden Atomic and Molecular Database (LAMD; \citealt {2005A&A...432..369S}) and Cologne Database for Molecular Spectroscopy (CDMS; \citealt{{2001A&A...370L..49M},{2005JMoSt.742..215M}}), assuming a gas temperature of 20~K.}
\label{spectra_detail}
\begin{tabular}{lcccl}
\hline      
Transition & Frequency ($\nu$) & $E_u/k_B$ & $n_{crit}$ & \multicolumn{1}{c}{Remarks} \\
& (GHz) & (K) & ($\rm cm^{-3}$) &           \\
\hline
$\rm C_2H\ (1-0)\ 3/2-1/2$ & 87.316925 & 4.19 & 2$\times 10^5$ & Tracer of photodissociation region \\
& & & & $\rm N_J = 1_{3/2} - 0_{1/2}$ is split into three hf components \\
$\rm HCN\ (1-0)$ & 88.631847 & 4.25 & 3$\times 10^6$ & tracer of high column density, optical depth \\
& & & & J = 1 - 0 is split into three hf components \\
$\rm HCO^+\ (1-0)$ & 89.188526 & 4.28 & 2$\times 10^5$ & High column density, kinematics \\
$\rm HNC\ (1-0)$ & 90.663572 & 4.35 & 3$\times 10^5$ & High column density, cold gas tracer \\
& & & & Three hf components \\ 
$\rm N_2H^+\ (1-0)$ & 93.173772 & 4.47 & 3$\times 10^5$ & High column density, depletion resistant, optical depth \\
& & & & J = 1 - 0 line has 15 hf components out of which seven have a different frequency \\
\hline
\end{tabular}
\end{table*}

\begin{table*}
\centering
\caption{\small Parameters of molecular transitions detected towards the two dust clumps. The LSR velocity ($\rm V_{LSR}$), line width ($\rm \Delta V$), main-beam temperature ($\rm T_{mb} $), and velocity integrated intensity ($\rm \int T_{mb} dV$) are obtained from the {\tt hfs} fitting method of CLASS90. 
The column density ($N_{\rm}$) values are estimated using RADEX and the fractional abundances ($x$) are derived using  mean $\rm H_2$ column density of $\rm 1.8 \times 10^{22}\ cm^{-2}$ and $\rm 1.0 \times 10^{22}\ cm^{-2}$ for Clump 3 and Clump 4, respectively (refer Section \ref{herschel}).}
\label{spectra_param}
\begin{tabular}{lcccccc}
\\ \hline 
Transition  & $\rm V_{LSR}$ & $\rm \Delta V$ & $\rm T_{mb} $ & $\rm \int T_{mb} dV$ & $N_{\rm}$ & $x$ \\
 & ($\rm km\ s^{-1}$) & ($\rm km\ s^{-1}$) & (K) & ($\rm K\ km\ s^{-1}$) & ($\rm \times 10^{14}\ cm^{-2}$) & ($10^{-9}$) \\
\hline      
\multicolumn{7}{c}{Clump 3} \\ 
\hline
$\rm C_2H$ & -30.10 & 2.61 & 0.27 & 0.33 & 3.59 & 19.90 \\
$\rm HCN$  & -30.80 & 3.31 & 0.38 & 1.25 & 6.05 & 33.63\\
$\rm HCO^+$ & -30.54 & 3.36 & 0.75 & 2.69 & 0.52 & 2.89 \\
$\rm HNC$  & -30.20 & 4.18 & 0.69 & 3.18 & 2.62 & 14.57 \\
$\rm N_2H^+$ & -29.50 & 2.38 & 0.77 & 2.46 & 3.11 & 17.29 \\
\hline 
\multicolumn{7}{c}{Clump 4} \\
\hline
$\rm C_2H$ & -32.40 & 1.96 & 0.36 & 0.86 & 2.47 & 23.97 \\
$\rm HCN$  & -32.40 & 1.90 & 0.31 & 0.76 & 1.71 & 16.64 \\
$\rm HCO^+$ & -32.71 & 4.33 & 0.52 & 2.38 & 0.28 & 2.75 \\
$\rm HNC$  & -32.50 & 3.18 & 0.38 & 1.50 & 0.69 & 6.70 \\
$\rm N_2H^+$ & -31.70 & 2.41 & 0.23 & 0.61 & 0.45 & 4.35 \\   
\hline   
\end{tabular}
\end{table*}

We have used RADEX \citep{2007A&A...468..627V}, a one-dimensional non-LTE radiative transfer code, to compute the column density of the detected molecular transitions. The input parameters to RADEX are the peak main beam temperature, back ground temperature (2.73~K from \citet{{2006MNRAS.367..553P},{2015MNRAS.451.2507Y}}), kinetic temperature, line width and $\rm H_2$ volume density. We assume the kinetic temperature to be same as dust temperature \citep{{2012ApJ...756...60S},{2016ApJ...833..248Y},{2016ApJ...818...95L}}. The $\rm H_2$ volume density and dust temperature values are taken 
from Section \ref{herschel} which are estimated using the {\it Herschel} maps. 
Using the mean $\rm N(H_2)$ for the clumps (listed in Table \ref{clump_properties}), we estimate the fractional abundances ($x$) of the detected molecules. The derived column densities and fractional abundances are listed in Table \ref{spectra_param}.
The values estimated for Clumps \#3 and \#4 are in the range derived for
a large sample of clumps associated with IRDCs in the work by
\citet{2014A&A...562A...3M}. Similar comparison holds with the results from \citet{2011A&A...527A..88V}. 

From Figure \ref{integrated}, it is evident that $\rm N_2H^+$ and HNC
molecular line emission is appreciably more extended towards Clump 3 in
comparison with Clump 4. This is supported by the derived column density values
which shows a clear decrease towards Clump 4. Further for $\rm N_2H^+$ the
fractional abundances and the integrated line intensities also decrease by more than a factor of 4 towards the central Clump 4. These indicate
that Clump 4 is possibly at a later evolutionary stage as compared Clump 3 \citep{{2014A&A...562A...3M},{2015MNRAS.451.2507Y},{2016ApJ...833..248Y}}. The radio peaks A and B fall within Clump 4 with B being close
to the peak in the {\it Herschel} column density map. Radio peak C is located
in Clump 3. The estimated $\rm C_2H$ and $\rm N_2H^+$ abundances show a decreasing 
trend with the increasing Lyman continuum photon flux of the clumps further suggesting  
an earlier evolutionary stage for Clump 3 \citep{{2016ApJ...833..248Y},{2015MNRAS.451.2507Y}}. Parameters derived from the other detected molecular transitions are fairly consistent with the above picture of evolutionary stage. 
However, it should be noted here that better signal-to-noise ratio
spectral observations are required before conclusively ascertaining the 
evolutionary stages of the two clumps.

\section{Feedback of high-mass stars and origin of CS51} \label{feedback}
Massive stars can influence the parental cloud via various feedback mechanisms.
In this section we attempt to understand this in connection with the origin
of the bubble CS51 and possible triggered star formation.

\subsection{IRAC band ratio images} \label{pah_des}
The general bubble structure is a PDR visible at 5.8 and 8~$\rm \mu m$ and an evacuated cavity within this 
\citep{{2006ApJ...649..759C},{2007ApJ...670..428C},{2008ApJ...681.1341W},{2009ApJ...694..546W},{2010A&A...518L..99A},{2010A&A...518L.101Z},{2010A&A...523A...6D},{2012ApJ...755...71K}}. The 5.8 and 8-$\rm \mu m$ emission is
attributed to PAH molecules, which are excited by the soft UV photons
permeating the PDR, with contribution from thermal emission from dust as well \citep{{2008ApJ...681.1341W},{2009A&A...494..987P}}. 
In their detailed study on M17, \citet{2007ApJ...660..346P}, proposed
the use of IRAC band ratio images to understand the interaction of massive
stars with their surrounding and delineate regions of PAH destruction.
They further confirmed the disappearance of PAH in the destruction zone from 
spectroscopic data. Following this pioneering work, \citet{2008ApJ...681.1341W}
used this technique in several bubbles to locate the PDR around them.
This technique exploits the fact that three of the four IRAC bands (at 3.6, 5.8, 8.0~$\rm \mu m$) sample several PAH emission features (see Table 5 of \citealt{2007ApJ...660..346P}) whereas the 4.5-$\rm \mu m$ band is PAH-free.

\begin{figure*}
\centering
\includegraphics[width=9cm,height=5cm,keepaspectratio]{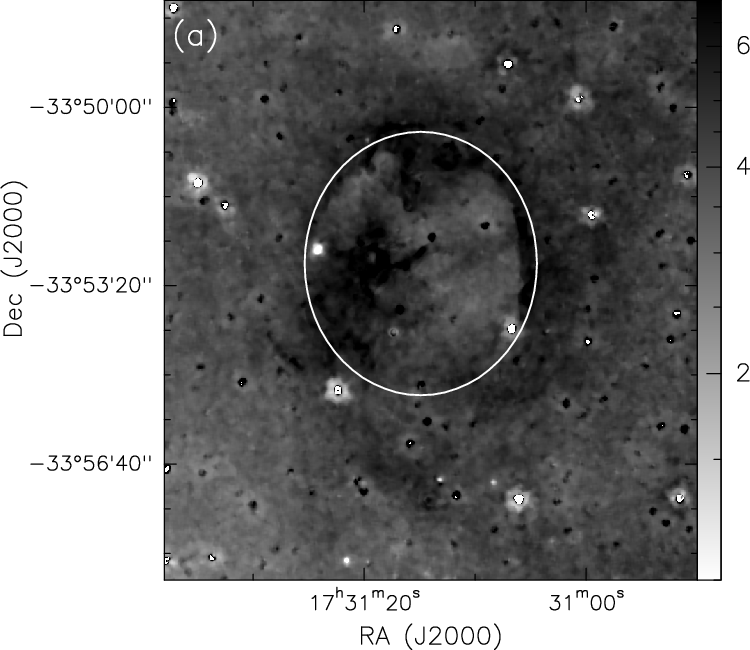}
\includegraphics[width=9cm,height=5cm,keepaspectratio]{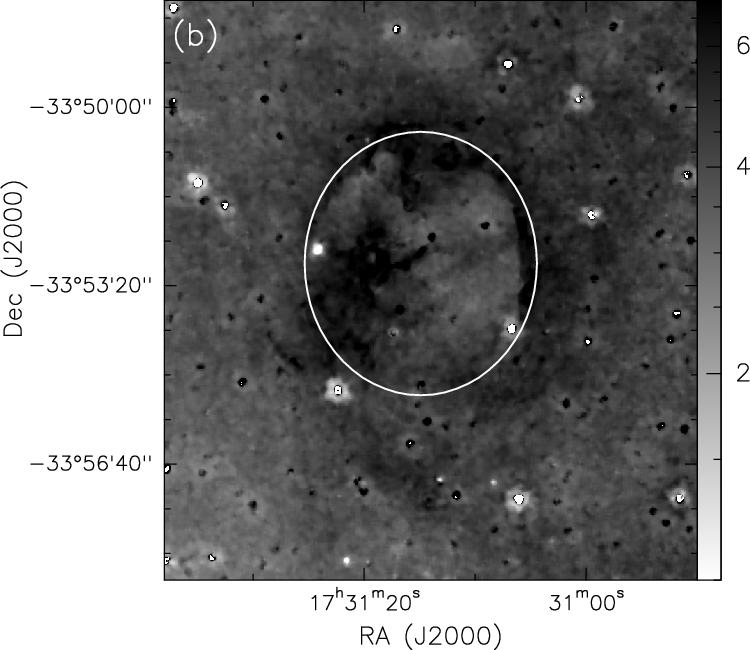}
\includegraphics[width=9cm,height=5cm,keepaspectratio]{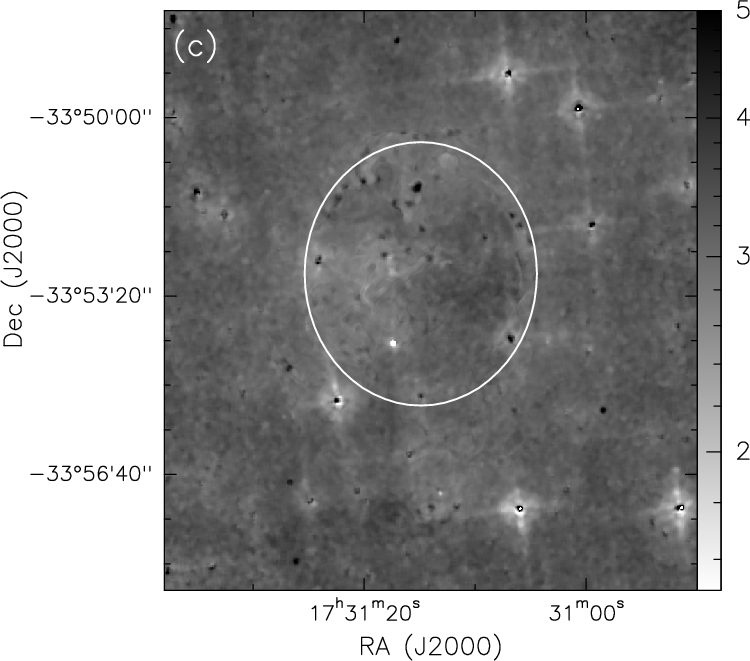}
\caption{IRAC ratio maps 8.0~$\rm \mu m$/4.5~$\rm \mu m$ (a), 5.8~$\rm \mu m$/4.5~$\rm \mu m$ (b) and 8.0~$\rm \mu m$/5.8~$\rm \mu m$ (c) of the region associated to bubble CS51. The bright emission in (a) and (b) show the PAH emission regions. The extent of the bubble is shown as a black ellipse.
Point source removal at 4.5~$\rm \mu m$ is not very good since the region is densely populated. The residuals show up in the ratio images (a) and (b). }
\label{pah_dest}%
\end{figure*}

We adopt the same procedure described in \citet{2007ApJ...660..346P} of using residual images (after removing point sources) followed by median filtering and
smoothening before taking the ratios.  
Figure \ref{pah_dest}, shows the ratio images. Figure \ref{pah_dest} (a) and (b)
show the ratio of 
8.0-$\rm \mu m$ and 5.8-$\rm \mu m$ images, respectively with the PAH-free 4.5-$\rm \mu m$ image. The bright emission (seen as dark in the colour inverted plots) towards the bubble rim in these ratio maps show the location of the PAH zones thus defining the PDR related to CS51. This is similar to the results obtained by \citet{2008ApJ...681.1341W} and \citet{2012ApJ...756..151D}. However, it should be noted that
the ratio maps also show enhanced emission towards the likely centre of the bubble
coinciding with the region of bright radio emission. Given that PAH would be
destroyed in the harsh radiation field close to the ionizing star (\citet{{2008ApJ...681.1341W},{2010A&A...523A...6D}} and references therein), this bright
region could possibly be from the thermal dust continuum which is supported
by the bright 24-$\rm \mu m$ emission that arises mostly near the hot star when the dust is heated to $\sim$100 K. Figure \ref{pah_dest} (c)
shows the ratio map between two PAH bearing bands (8.0~$\rm \mu m$/5.8~$\rm \mu m$). As expected, the rim emission is invisible here.
It is to be noted that while these ratio images give the overall picture of
PAH emission in the PDR, it should be kept in mind that the contribution from thermal dust and other atomic and molecular features would also be present. Thus
these ratio maps should be taken as illustrative unless confirmed with spectroscopy.  
Nevertheless, the ratio maps do confirm with the generally accepted picture
of bubbles. 

\subsection{Column density probability distribution function} \label{prob_den}
Structures in the form of clumps or condensations at the border of IR dust bubbles,
pillars and protrusions pointing towards the ionizing source have been observed in several studies \citep{{2009A&A...494..987P},{2010A&A...513A..44P},{2010A&A...518L.101Z},{2012A&A...542A..10A},{2012A&A...544A..39J},{2015ApJ...798...30L},{2016ApJ...818...95L}}. The formation of these structures and the star formation therein are discussed invoking the `collect
and collapse' (CC) model \citep{1977ApJ...214..725E} or the `radiation driven implosion' (RDI) model \citep{{1989ApJ...346..735B},{1994A&A...289..559L}}, thus probing
the feasibility of triggered star formation in the `accumulated' versus `pre-existing' scenarios.

Understanding the role of density compression of ionized gas into a molecular cloud
and that of ram pressure of turbulence of the cloud are crucial aspects in 
deciphering the impact of high-mass stars and this is discussed in detail by \citet{{2012A&A...546A..33T},{2012A&A...538A..31T},{2014A&A...564A.106T}}. They use the probability distribution
function (PDF) of the column density around the ionized regions to study
the feedback of the massive star. Their studies show that the column density PDF
displays a single peak if turbulence dominates and the distribution
becomes bimodal with a second peak forming at higher densities if the 
ionization pressure becomes larger than the turbulence. 
\citet{{2016ApJ...818...95L},{2017A&A...602A..95L}} have implemented this technique on two \hii\ regions / bubbles. 
Adopting the same formulation, we investigate the nature of the column density
PDFs in the region associated with CS51. 
We generate the PDF for the regions encircled within the three circles
shown in Figure \ref{dust_temp}(a). These concentric circles are constructed
with a separation of 1\arcmin~and for clarity we name them as Circle 1, 2, and 3
in increasing order of radius. 
The functional form for the bimodal PDF used to fit the column density is given by the following expression \citep{{2016ApJ...818...95L},{2012A&A...540L..11S},{2014A&A...564A.106T}}
\begin{equation}
\begin{split}
p(\eta) = \frac{p_0}{\sqrt{2\pi\sigma_0^{2}}}exp\left(\frac{-(\eta-\mu_0)^2}{2\sigma_0^2}\right)
+\frac{p_1}{\sqrt{2\pi\sigma_1^{2}}}exp\left(\frac{-(\eta-\mu_1)^2}{2\sigma_1^2}\right)
\end{split}
\end{equation}
 where, $\eta={\rm ln}(N/\bar{N})$, $N$ is the column density and $\bar{N}$ is the average column density taken over Circle 3. $p_i$, $\mu_i$, and $\sigma_i$ are the integral, mean, and dispersion of each component. In the above expression, the first lognormal component at low column density is associated with the initial turbulent
molecular cloud and the second component at high column density is attributed to compression by the ionized gas pressure \citep{{2016ApJ...818...95L},{2014A&A...564A.106T}}. 

\begin{figure}
\centering
\includegraphics[scale=0.3]{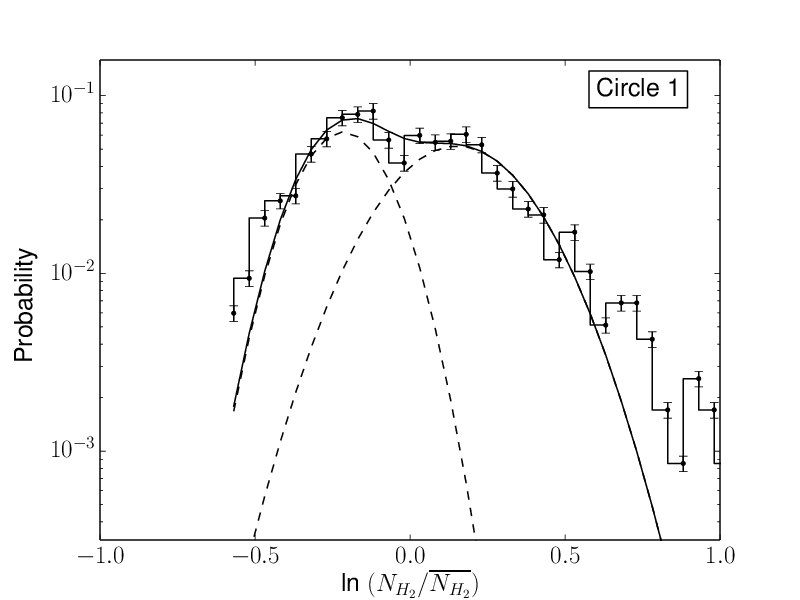}
\includegraphics[scale=0.3]{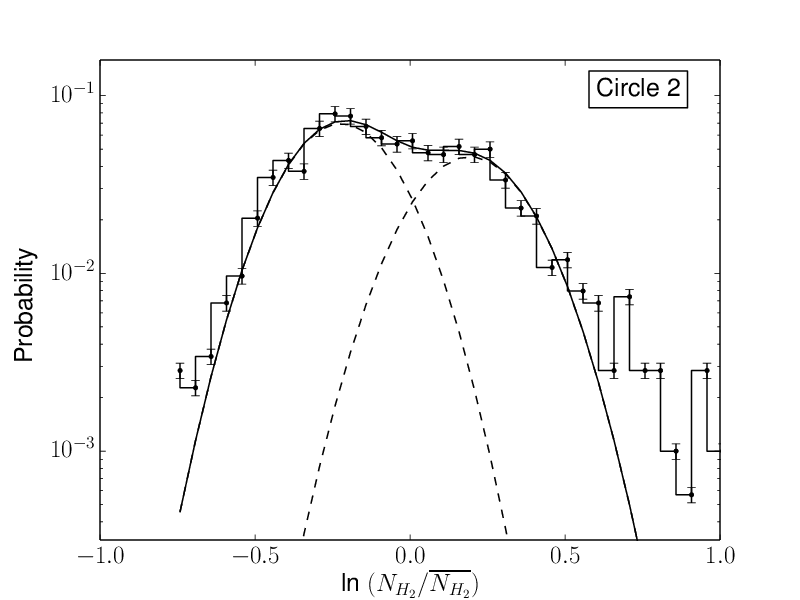}
\includegraphics[scale=0.3]{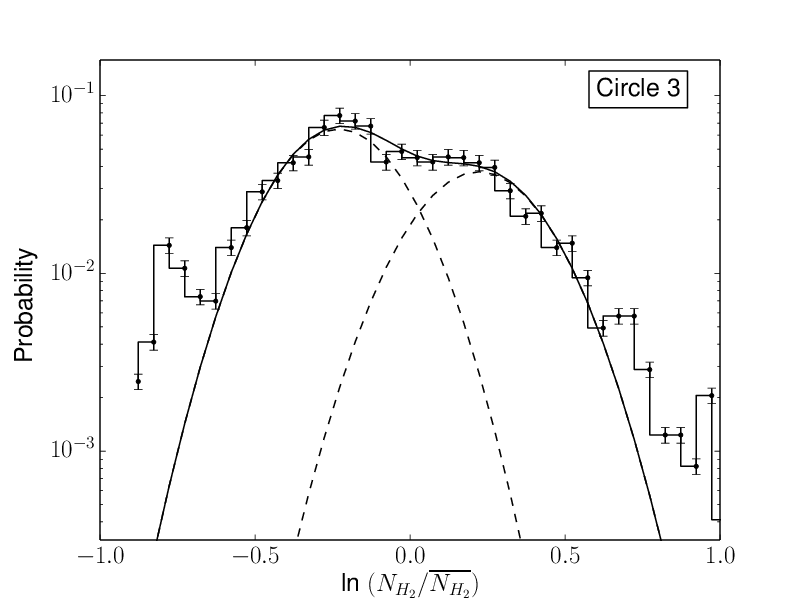}
\caption{Plots of column density PDFs over the three circular regions. The dashed lines are the two lognormal components and the best fit (black solid line) is the sum of these two. $\overline{N_{H_2}}$ is the mean column density computed over the largest region. }
\label{PDF}%
\end{figure}

Figure \ref{PDF} shows the column density PDFs
in the three identified circles covering the ionized emission associated
with CS51 and the surrounding cloud. The fitted parameters are listed in 
Table \ref{pdf_parm}. The PDFs fit fairly well and all of them clearly show the
second peak consistent with the results obtained for the bubbles N4, RCW 79
and RCW 120 \citep{{2014A&A...564A.106T},{2016ApJ...818...95L},{2017A&A...602A..95L}}. Similar trend
of $p_0$ increasing and $p_1$ decreasing as we move outwards is seen for
CS51. As noted by \citet{2016ApJ...818...95L} and discussed by \citet{2014A&A...564A.106T}, the decrease in $p_1$ with increasing radius indicates decreasing effect
of compression due to ionized gas for larger regions.
The results obtained reiterates the strong influence of the expanding bubble on its immediate
surrounding. The second lognormal form suggesting compression from ionized gas 
would likely account for the condition required for triggered star formation. The high column density region shows signature of a power-law tail, which is generally attributed to on-going star formation \citep{{2015A&A...575A..79S},{2012A&A...540L..11S},{2017A&A...602A..95L},{2013ApJ...766L..17S},{2013A&A...554A..42R}}. Insufficient data points and low signal-to-noise ratio prevents us from
fitting a power-law and attempting a more quantitative comparison with the above
studies.

\begin{table*}
\centering
\caption{Retrieved parameters from the lognormal fit to column density PDFs.}
\label{pdf_parm}
\small
\begin{tabular}{cccccccc}
\\ \hline
Circle & radius ($\arcmin$)&$p_0$ & $\mu_0$ & $\sigma_0$ & $p_1$ & $\mu_1$ & $\sigma_1$ \\
\hline
1 & 4.5 & 0.020 & -0.215 & 0.131 & 0.027 & 0.152 & 0.206  \\
2 & 5.5 & 0.028 & -0.222 & 0.163 & 0.019 & 0.193 & 0.171 \\
3 & 6.5 & 0.029 & -0.229 & 0.179 & 0.018 & 0.222 & 0.190  \\
\hline                
\end{tabular}
\end{table*}

\subsection{Collect and collapse scenario}\label{CC}
The previous sub-section showed the role of ionization compression in CS51. The
effect of this in triggered star formation has been in focus since the pioneering
work by \citet{1977ApJ...214..725E}. As mentioned earlier, several studies
have focused on the mechanism behind this triggering process - CC or RDI. 
The detection of five dense clumps in the PDR region suggest that the
shocked molecular layer is mostly swept up and accumulated during the
bubble expansion. Thus we explore the possibility of the CC model at work
around CS51. To examine this, we estimate the dynamical age of the bubble and
compare the same with the fragmentation time of the surrounding cloud.

Assuming that the \hiir\ expands in a homogeneous molecular cloud, the dynamical age of the \hiir\ is estimated from the following well known expressions \citep{{1978ppim.book.....S},{1980pim..book.....D}}

\begin{equation}
\rm R_{st} = \left[\frac{3\ N_{Lyc}}{4\ \pi\ n_{0}^{2}\ \alpha_{B}} \right]^{1/3}
\label{stromgren}
\end{equation}

\begin{equation}
\rm t_{dyn} = \frac{4}{7}\ \frac{R_{st}}{C_{Hii}} \left[ \left(\frac{R_{if}}{R_{st}}\right)^{7/4}\ -1\right]
\end{equation}
where, $\rm R_{st}$ is the Str\"omgren radius, $\rm N_{Lyc}$ is the Lyman continuum photons flux responsible for the \hiir\ and $\rm n_{0}$ is the initial particle density of the ambient gas. $\rm \alpha_{B}$ is the coefficient of radiative recombination and it is calculated from \citet{1997ApJ...489..284K}, to be 2.6$\times10^{-13}$ $\rm (10^{4}\ K/T)^{0.7}$ $\rm cm^{3}\ sec^{-1}$.
$\rm t_{dyn}$ is the dynamical age of the \hiir,\ $\rm C_{Hii}$ is the isothermal sound speed (assumed to be 10~$\rm km\ s^{-1}$), $\rm R_{if}$ is the radius of \hiir\ which is approximately taken as the radius of the bubble (2.1$\arcmin$). To estimate $\rm n_{0}$, we take the 
total mass of the swept up molecular shell which is the sum of the masses
of the five detected clumps on the bubble periphery (11630~$\rm M_{\odot}$)
and assume that it was distributed homogeneously within the bubble radius. 
To this we add the mass of the ionized component (300~$\rm M_{\odot}$).
This
gives an estimate of $\rm n_{0} = 2.4\times 10^3\ cm^{-3}$ which defines
a lower limit since it does not take into account the ionized gas
that probably would have escaped through the rupture of bubble or absorbed
by dust. From the estimated parameters of the clumps (see Table \ref{clump_properties}), we get an average number density of  $\rm n_{0} = 4.8\times 10^3\ cm^{-3}$ which can be considered as the upper limit.  
Using the
above values, we estimate the dynamical age of CS51 to be 0.9 -- 1.3~Myr. This estimate should be taken with caution since the assumption of
expansion in a uniform medium is not realistic. 

We proceed further to investigate fragmentation of the collected shell 
material using the model of \citet{1994MNRAS.268..291W}. They give the
following expression to estimate the fragmentation time.

\begin{equation}
\rm t_{frag} = 1.56\ a_{s}^{7/11}\ N_{49}^{-1/11}\ n_3^{-5/11}\ Myr
\end{equation}
where, $a_{s}$ is the sound speed in the shocked layer (the turbulent velocity) in units of 0.2~$\rm km\ sec^{-1}$, $\rm N_{49}$ is the ionizing photon in units of $10^{49}\ \rm photons\ sec^{-1}$, and  $\rm n_3$ is the initial particle density 
of the ambient gas ($\rm n_{0}$) in units of $10^3~ \rm cm^{-3}$. 
Taking $a_{s}$ to be 0.3~$\rm km\ s^{-1}$ at the derived dust temperatures,
we estimate the fragmentation time, $\rm t_{frag}$, to be 0.5~Myr for $\rm n_{0} = 2.4\times 10^3\ cm^{-3}$ and 0.4~Myr for $\rm n_{0} = 4.8\times 10^3\ cm^{-3}$. The estimated fragmentation time is shorter 
(by a factor of 2) than the dynamical
age of the \hiir\ implying that the shell of collected material has enough
time to gravitationally fragment during the expansion of the bubble. Similar
results are obtained for N4, G8.14+0.23 and G24.136+00.436, where the fragmentation time is less than the dynamical time \citep{{2016ApJ...818...95L},{2015ApJ...798...30L},{2012ApJ...756..151D}}. 
In contrast, 
\citet{2012A&A...544A..39J} derive fragmentation time significantly larger
than the dynamical age for bubble N22 and they discard the CC process. The above estimates of the timescales involved support the CC process as the likely 
mechanism for the formation of the identified clumps and the associated star formation.
 It should however be kept in mind that, this simple model does not
conclusively exclude the
RDI process and thus higher resolution observations are required to probe the
dense cores within these clumps and compare their masses, radii and separation with the theoretical predictions of the CC mechanism \citep{{1994MNRAS.268..291W},{2017arXiv170504907L}}.

\subsection{What do the clumps reveal?} \label{clump_starformation}
As discussed in Section \ref{mol_emm}, the Clumps \#3 and \#4 show trends of
different evolutionary stages with Clump 4 being more evolved. Association with
radio peaks and 24~$\rm \mu m$ emission indicate that these are active star forming clumps. Further, the NIR and MIR colours have shown the presence of YSOs likely to
be associated with the clumps with an overdensity of Class I YSOs seen towards
Clump 3. 

From the derived centroid velocities of the molecular transitions in the clumps,
it is seen that the lines related to Clump 3 are red-shifted by
$\sim 2~ \rm \, km\ s^{-1}$ with respect
to a systematic velocity of -32.34 $\rm km\ s^{-1}$ which is taken as the average of velocities of detected molecular line emission in Clump 4. Clump 4 is located within the bubble (and closer to the centre) correlating well with the likely location of the
exciting star(s). The above is consistent with the picture of the formation
of CS51 where the shell of dust and gas seen is presumably swept up by the
expanding \hii\ region. Hence Clump 3 that is located towards the northern rim
is expected to be moving away from the central clump indicating the expansion
of the bubble. Similar expansion is revealed from the molecular line data of
bubble N4 where the clumps towards the south-east and north-west part of N4 are
seen to have blue-shifted and red-shifted velocities, respectively compared to
a systematic velocity. It should be noted here that we do not have molecular line
data for Clumps \#1, \#2, and \# 5 to strengthen the picture of expansion.

From the mass and size estimates of the five clumps discussed in Section \ref{herschel}, Clumps 2, 3, 4, and 5 qualify as potential high-mass forming ones as they satisfy the threshold criteria of $\rm m(r)\geqslant 870\ M_{\odot}\ (r/pc)^{1.33}$, where m(r) is mass of clump and r is the effective radius of clump, discussed in \citet{2010ApJ...723L...7K}.   
We have also investigated the gravitational state of Clumps \#3 and \#4 by deriving
the virial parameter ($\alpha$) defined as the ratio between virial mass to dust mass. This parameter indicates whether a clump is gravitationally bound or not.
Clumps are likely to undergo gravitational collapse if $\alpha$ $<$ 1 and in
the absence of magnetic field can be considered as bound. If the value of $\alpha$
exceeds unity, then in all likelihood expansion is possible unless some
external mechanism constrains the cloud.
Virial mass of each clump is derived from the following equation \citep{2016MNRAS.456.2041C}. 
\begin{equation}
\rm M_{vir} = \frac{5\ r\ \Delta V^2}{8\ ln(2)\ a_1\ a_2\ G} \sim 209\ \frac{1}{a_1\ a_2} \left(\frac{\Delta V}{km\ s^{-1}} \right)^2\ \left(\frac{r}{pc}\right) M_{\odot}
\end{equation}
where, the constant $\rm a_1$ is the correction for power-law distribution, which can be expressed as $\rm a_1$ = (1-p/3)/(1-2p/5), for p $<$ 2.5 \citep{1992ApJ...395..140B}. We adopt a power-law density distribution of p=1.8 \citep{2016MNRAS.456.2041C}. The constant $\rm a_2$ is correction for non-spherical shape. We assume the clumps to be spherical and take $\rm a_2$ as 1. $\rm \Delta V$ and r are the line width and radius (listed in Table \ref{clump_properties}) of the clumps. To derive virial mass and virial parameter, we use the optically thin $\rm N_2H^+$ emission because an optically thick line would overestimate the virial mass  \citep{2012ApJ...756...60S}. 
Thus using the above equation and the line width obtained in Section \ref{mol_emm}, we derive virial masses of $\sim$ 1100 and 835 $\rm M_{\odot}$ for
Clumps 3 and 4, respectively. Taking the ratio with the mean dust masses of the
clumps (see Table \ref{clump_properties}), we estimate the virial parameter to
be 0.2 and 0.6 for Clumps 3 and 4, respectively. These estimates are consistent
with the nature of the clumps where signature of active star formation is observed. 

\section{Summary} \label{summary}
In this work, we have carried out a detailed multi-wavelength study towards the southern IR bubble CS51 which is associated with IRAS 17279-3350. Our main results are the following.
\begin{enumerate}

\item The associated ionized emission probed at 610 and 1300~MHz displays a complex morphology. The ionized emission mostly fills the bubble interior with a cavity towards the south-west, which is more pronounced at 1300~MHz. 
The observed Lyman continuum photon flux translates to a spectral type of O6V -- O5.5V for the exciting source under the assumption of optically thin, free-free emission. The mass of this associated ionized gas within the bubble is 
estimated to be $\sim\ 300 \rm \, M_{\odot}$.

\item The GMRT radio maps reveal the presence of three components (A, B, and C) associated with CS51. They show up as compact regions of enhanced emission. The component A is located towards the likely center of the bubble, the component B (brightest among them) is towards the south-east periphery and the component C is located in the north-west rim of the bubble. 

\item The 610 - 1300~MHz spectral index map shows the coexistence of both thermal free-free emission and non-thermal synchrotron emission. The compact regions A and C are seen to be associated with thermal emission and component B shows signature of non-thermal emission.

\item From the optical (B-band) and NIR colours, we identify three sources as
potential candidate exciting star(s) from a sample of sources located towards
the radio and 24~$\rm \mu m$ emission peaks.   

\item The column density and dust temperature maps are generated by pixel-wise SED modeling of the thermal dust emission using FIR {\it Herschel} data. The column density map shows a fragmented shell like structure harbouring four clumps towards
the periphery of the bubble and one clump is detected within the bubble close to the likely centre. Dust temperature map shows the presence of warmer dust in the bubble interior
consistent with the presence of ionized emission. 

\item Five molecular transitions ($\rm C_2H$, HCN, $\rm HCO^+$, HNC, and $\rm N_2H^+$) are detected towards two clumps (\#3 and \#4). The hyperfine components of 
$\rm C_2H$, HCN, HNC, and $\rm N_2H^+$ are clearly detected above noise level.
The velocity integrated (0th moment) map for $\rm N_2H^+$ and HNC are seen to be
appreciably extended towards Clump 3. Within the uncertainties due to low signal-to-noise ratio of the observed spectra, the derived line parameters, column density,
and fractional abundance suggest a possible earlier evolutionary phase of Clump 3
as compared to Clump 4. 

\item The IRAC ratio maps involving PAH-bearing and PAH-free bands show the
presence of PAH emission associated with the PDR of CS51 thus confirming the
generally accepted bubble formation mechanism and related MIR emission. 

\item The column density PDFs display a bimodal distribution thus demonstrating the strong influence of expanding bubble on its surrounding where compression due to
ionized gas pressure dominates the effect of turbulence. 

\item Assuming the expansion of the associated \hii\ region to occur in a uniform medium, the dynamical age is estimated to be 0.9 -- 1.3~Myr, which is higher than the derived fragmentation time of 0.4 -- 0.5~Myr. This indicates that the collect and collapse mechanism is possibly at work around the bubble CS51 and
responsible for triggering star formation towards the border of the bubble.

\item The estimated values of mass and radius of the clumps, that lie between 810 -- 4600$\rm M_{\odot}$ and 1.0 -- 1.9~pc, respectively, indicate all except Clump 1 to be high-mass star forming clumps.  
 
\item The centroid velocities of the molecular transitions detected show signatures
of expansion of the bubble. The Clump 3 located towards the northern rim is seen
to have red-shifted velocities with respect the central Clump 4.

\end{enumerate}

\begin{small}
\textit{Acknowledgment : We place on record our sincere thanks to the referee
for carefully going through the manuscript and giving valuable suggestions. We thank the staff of the GMRT, that made the radio observations possible. GMRT is run by the National Centre for Radio Astrophysics of the Tata Institute of Fundamental Research. This work is based [in part] on observations made with the {\it Spitzer} Space Telescope, which is operated by the Jet Propulsion Laboratory, California Institute of Technology under a contract with NASA. This publication made use of data products from {\it Herschel} (ESA space observatory) and the Millimetre Astronomy Legacy Team 90 GHz (MALT90) survey. }
\end{small}

\bibliography{refer}

\begin{appendix}
\section{YSO list.}

\begin{table*}
\centering
\tiny
\caption{\small List of identified YSOs detected in region of bubble CS51. The 2MASS and IRAC photometric magnitudes of the YSOs are given.}
\label{yso_table}
\begin{tabular}{ccccccccccccccc}
\\ \hline 
YSO & RA (J2000) & DEC (J2000) & J & H & K & 3.6 & 4.5 & 5.8 & 8.0 & 24 & \multicolumn{3}{c}{YSO Class} \\
 &$(^h~~^m~~~^s)$ & (~$^\circ~~\arcmin~~~\arcsec$)&(mag)&(mag)&(mag)&(mag) &(mag)&(mag)&(mag)&(mag)& 1 & 2 & 3 & 4\\
 \hline 
1 & 17 30 51.95 & -33 52 21.04 & -- & -- & -- & 12.88 & 12.46 & 11.87 & 11.42 & -- & Class I/II & -- & Class II & --\\
2 & 17 30 53.83 & -33 54 39.49 & -- & -- & -- & 12.70 & 12.62 & 11.82 & 11.97 & -- & -- & -- & Class II & -- \\
3 & 17 30 54.05 & -33 51 14.72 & 14.35 & 12.74 & 12.07 & 11.49 & 11.32 & 11.27 & 10.83 & -- & Class II & -- & -- & --\\
4 & 17 30 54.77 & -33 50 58.34 & 15.27 & 13.33 & 12.70 & 11.86 & 11.82 & 12.17 & 11.75 & -- & Class II & -- & -- & --\\
5 & 17 30 55.05 & -33 50 42.22 & -- & -- & -- & 13.21 & 13.25 & -- & -- & 7.93 & -- & Class II & -- & --\\
6 & 17 30 56.78 & -33 54 13.90 & 13.44 & 11.25 & 10.32 & 9.68 & 9.71 & 9.48 & 9.42 & 6.24 & -- & Class II & -- & --\\ 
7 & 17 30 58.02 & -33 54 17.63 & 15.11 & 13.41 & 12.42 & 11.54 & 11.46 & 11.32 & -- & -- & -- & -- & -- & Class II  \\
8 & 17 30 58.15 & -33 50 39.01 & -- & -- & -- & 11.55 & 10.97 & 10.67 & 10.55 & -- & -- & -- & Class II & --\\
9 & 17 30 58.38 & -33 55 00.12 & -- & -- & -- & 11.83 & 11.67 & 12.22 & 11.57 & -- & Class II & -- & -- & -- \\
10 & 17 30 59.46 & -33 52 01.49 & 12.84 & 10.51 & 8.74 & 6.70 & -- & 4.79 & 4.21 & 3.36 & -- & -- & -- & Class I\\
11 & 17 30 59.52 & -33 54 09.25 & 14.90 & 12.88 & 12.05 & 11.38 & 11.38 & 11.12 & 10.63 & -- & -- & -- & Class II & -- \\
12 & 17 31 01.19 & -33 51 07.09 & 13.24 & 10.43 & 9.17 & 8.24 & 8.25 & 7.92 & 7.86 & 5.02 & -- & Class II & -- & -- \\

13 & 17 31 01.35 & -33 50 53.81 & -- & -- & -- & 13.22 & 12.60 & -- & -- & 3.10 & -- & Class I & -- & --\\ 
14 & 17 31 01.35 & -33 51 21.95 & 14.07 & 13.05 & 12.46 & -- & -- & -- & -- & -- & -- & -- & -- & Class II \\
15 & 17 31 01.97 & -33 54 16.62 & 14.57 & 12.58 & 11.23 & 10.19 & 10.05 & 9.71 & 9.82 & -- & -- & -- & -- & Class II \\
16 & 17 31 02.12 & -33 53 32.32 & -- & -- & 12.01 & 11.21 & 11.30 & 10.84 & 10.23 & -- & -- & -- & Class II & --\\
17 & 17 31 03.24 & -33 53 58.38 & 14.74 & 12.41 & 11.53 & 10.80 & 10.84 & 10.47 & 10.11 & -- & -- & -- & Class II & --\\
18 & 17 31 03.37 & -33 53 53.77 & 14.52 & 12.44 & 11.68 & 11.06 & 11.08 & 10.58 & 10.32 & -- & -- & -- & Class II & --\\
19 & 17 31 03.79 & -33 51 48.56 & -- & -- & 11.64 & 9.66 & 9.06 & 8.56 & 8.68 & -- & -- & -- & Class II & --\\
20 & 17 31 03.87 & -33 49 04.58 & -- & -- & -- & 10.92 & 9.52 & 8.46 & 7.78 & 3.94 & Class I & Class I & Class I & --\\
21 & 17 31 03.91 & -33 50 51.61 & 15.18 & 13.02 & 12.28 & 11.60 & 11.55 & 11.37 & 10.84 & -- & Class II & -- & Class II & --\\
22 & 17 31 03.94 & -33 50 29.04 & -- & -- & -- & 11.79 & 11.83 & 11.55 & 10.89 & -- & -- & -- & Class II & --\\
23 & 17 31 03.96 & -33 48 34.63 & -- & -- & -- & 13.98 & 13.31 & 11.42 & 10.62 & -- & Class I/II & -- & Class I & --\\
24 & 17 31 04.63 & -33 54 44.86 & -- & -- & -- & 13.30 & 13.28 & 11.81 & 10.33 & -- & -- & -- & Class I & --\\
25 & 17 31 05.47 & -33 53 21.16 & -- & -- & -- & 12.41 & 12.34 & -- & -- & 4.15 & -- & Class I & -- & --\\
26 & 17 31 06.01 & -33 50 57.52 & 15.00 & 12.96 & 12.13 & 11.36 & 11.46 & 11.09 & 10.57 & -- & -- & -- & Class II & --\\
27 & 17 31 06.37 & -33 51 33.52 & 14.74 & 12.66 & 11.79 & 11.23 & 11.20 & 10.43 & 9.30 & -- & -- & -- & Class II & --\\
28 & 17 31 06.46 & -33 55 28.06 & -- & 10.68 & 9.84 & 9.34 & 9.43 & 9.04 & -- & 5.06 & -- & Class II & -- & --\\
29 & 17 31 07.16 & -33 50 29.33 & 13.53 & 10.85 & 9.59 & 8.74 & 8.62 & 8.27 & 8.08 & 5.13 & -- & Class II & -- & --\\
30 & 17 31 07.17 & -33 52 45.98 & 13.66 & 12.89 & 12.39 & 12.22 & 12.13 & 12.06 & -- & -- & -- & -- & -- & Class II \\
31 & 17 31 07.57 & -33 52 44.69 & 14.47 & 13.55 & 13.28 & 12.39 & 12.14 & 12.06 & 11.10 & -- & Class II & -- & Class II & -- \\
32 & 17 31 07.80 & -33 54 58.28 & 12.80 & 10.59 & 9.48 & 8.94 & 8.78 & 8.27 & 8.04 & -- & -- & -- & Class II & --\\
33 & 17 31 07.94 & -33 49 14.61 & 15.24 & 13.53 & 12.57 & 11.56 & -- & 11.62 & -- & -- & -- & -- & -- & Class II \\
34 & 17 31 08.08 & -33 50 56.72 & -- & -- & -- & 8.12 & 7.62 & 7.20 & 7.10 & -- & -- & -- & Class II & -- \\
35 & 17 31 08.22 & -33 48 34.77 & 14.87 & 13.54 & 12.69 & 11.99 & 12.06 & 12.14 & -- & -- & -- & -- & -- & Class II \\
36 & 17 31 08.91 & -33 48 47.99 & -- & 13.73 & 12.70 & 11.87 & 11.98 & 11.78 & 11.15 & -- & -- & -- & Class II & --\\
37 & 17 31 09.17 & -33 54 06.02 & 14.17 & 12.35 & 11.29 & 10.21 & 10.05 & 9.81 & 9.89 & -- & -- & -- & -- & Class II \\
38 & 17 31 09.33 & -33 48 53.77 & 12.61 & 11.71 & 11.16 & 10.64 & 10.67 & 10.42 & 10.61 & -- & -- & -- & -- & Class II \\

39 & 17 31 09.34 & -33 56 03.37 & -- & -- & -- & 14.01 & 14.06 & -- & -- & 3.38 & -- & Class I & -- & --\\
40 & 17 31 09.61 & -33 54 01.04 & -- & -- & -- & 13.15 & 12.01 & 10.92 & 10.19 & -- & Class I & -- & Class I & --\\
41 & 17 31 09.67 & -33 55 20.24 & -- & 13.23 & 12.35 & 11.67 & 11.59 & 11.38 & 10.97 & -- & Class II & -- & -- & --\\
42 & 17 31 09.87 & -33 53 02.93 & 14.38 & 13.35 & 12.73 & 12.23 & 12.05 & -- & -- & -- & -- & -- & -- & Class II \\
43 & 17 31 10.38 & -33 56 40.92 & 14.91 & 12.73 & 11.95 & 11.28 & 11.26 & 10.66 & 9.95 & -- & Class II & -- & Class II & --\\
44 & 17 31 10.59 & -33 49 00.66 & -- & -- & -- & 13.15 & 13.10 & 11.71 & 10.13 & -- & -- & -- & Class I & --\\
45 & 17 31 10.67 & -33 56 42.07 & 14.82 & 12.89 & 12.08 & 11.34 & 11.39 & 10.67 & 9.94 & -- & -- & -- & Class II & --\\
46 & 17 31 12.25 & -33 48 38.81 & 11.64 & 9.34 & 8.19 & 7.27 & 7.15 & 6.86 & 6.76 & 4.16 & -- & Class II & -- & --\\
47 & 17 31 12.44 & -33 50 47.47 & -- & -- & -- & 8.78 & 8.55 & 8.12 & 8.16 & 3.14 & -- & Class I/II & -- & --\\
48 & 17 31 12.84 & -33 49 05.16 & -- & -- & -- & 11.78 & 11.79 & -- & -- & 3.18 & -- & Class I & -- & --\\
49 & 17 31 13.04 & -33 50 13.99 & 11.97 & 10.64 & 10.09 & 9.72 & 9.85 & 9.52 & 9.00 & -- & -- & -- & Class II & --\\
50 & 17 31 14.78 & -33 51 55.48 & -- & -- & -- & 13.12 & 12.43 & 10.26 & 8.82 & -- & Class I & -- & Class I & --\\
51 & 17 31 14.84 & -33 48 51.71 & 14.92 & 13.27 & 12.346 & 11.66 & -- & 11.53 & -- & -- & -- & -- & -- & Class II \\
52 & 17 31 15.43 & -33 56 33.73 & 13.71 & 12.33 & 11.57 & -- & -- & -- & -- & -- & -- & -- & -- & Class II \\
53 & 17 31 15.71 & -33 50 39.05 & -- & -- & -- & 13.17 & 13.03 & 10.86 & 9.76 & -- & -- & -- & Class I & --\\
54 & 17 31 16.18 & -33 54 17.09 & 15.06 & 13.62 & 12.80 & 11.85 & -- & -- & -- & -- & -- & -- & -- & Class II \\
55 & 17 31 16.38 & -33 53 08.67 & 12.67 & 11.56 & 10.93 & -- & -- & -- & -- & -- & -- & -- & -- & Class II \\
56 & 17 31 17.28 & -33 51 35.39 & -- & -- & -- & 11.96 & 11.47 & 9.21 & -- & 1.44 & -- & Class I & -- & --\\
57 & 17 31 17.48 & -33 52 55.47 & 11.73 & 10.51 &  9.79 & -- & -- & -- & -- & -- & -- & -- & -- & Class II \\
58 & 17 31 18.85 & -33 55 31.94 & -- & -- & -- & 13.36 & 13.23 & -- & -- & 5.98 & -- & Class I & -- & --\\
59 & 17 31 18.86 & -33 57 30.69 & -- & -- & -- & 13.15 & 13.39 & -- & -- & 6.78 & -- & Class I/II & -- & --\\
60 & 17 31 19.45 & -33 52 13.80 & -- & -- & -- & 12.59 & 11.27 & 9.18 & 7.75 & -- & Class I & -- & Class I & --\\
61 & 17 31 19.58 & -33 51 14.40 & -- & 13.45 & 12.56 & 11.82 & 11.79 & -- & -- & 0.66 & -- & Class I & -- & --\\
62 & 17 31 19.66 & -33 54 39.05 & 13.33 & 11.68 & 10.76 & -- & -- & -- & -- & -- & -- & -- & -- & Class II \\
63 & 17 31 20.14 & -33 48 44.73 & 11.38 & 10.26 & 9.56 & -- & -- & -- & -- & -- & -- & -- & -- & Class II \\
64 & 17 31 20.26 & -33 57 16.66 & 14.56 & 12.94 & 12.00 & 11.51 & -- & 11.38 & -- & -- & -- & -- & -- & Class II \\
65 & 17 31 20.39 & -33 54 21.39 & 14.51 & 13.09 & 11.68 & 10.51 & 10.28 & 9.74 & -- & -- & -- & -- & -- & Class I \\
66 & 17 31 21.26 & -33 50 37.28 & 12.56 & 10.34 & 9.24 & 8.19 & 7.81 & 7.15 & 6.38 & 4.10 & Class I/II & Class II & Class II & --\\
67 & 17 31 21.38 & -33 57 28.76 & -- & -- & -- & 11.73 & 11.69 & 11.38 & 10.99 & -- & Class II & -- & Class II & --\\
68 & 17 31 21.57 & -33 51 09.78 & 13.61 & 11.73 & 10.69 & -- & -- & -- & -- & -- & -- & -- & -- & Class II \\
69 & 17 31 22.86 & -33 55 28.49 & 14.99 & 12.91 & 11.68 & -- & -- & -- & -- & -- & -- & -- & -- & Class II \\
70 & 17 31 23.48 & -33 54 47.12 & -- & 13.01 & 12.20 & 11.50 & 11.62 & 10.63 & 9.45 & -- & -- & -- & Class II & --\\
71 & 17 31 24.40 & -33 56 29.67 & 14.17 & 12.34 & 11.29 & -- & -- & -- & -- & -- & -- & -- & -- & Class II \\
72 & 17 31 24.41 & -33 55 58.66 & -- & -- & 12.54 & 11.37 & 11.19 & 10.65 & 10.56 & -- & -- & -- & Class II & --\\
73 & 17 31 25.33 & -33 49 55.31 & 13.29 & 12.56 & 12.19 & 11.84 & 11.45 & 11.59 & 11.29 & -- & Class I & -- & -- & --\\
74 & 17 31 26.66 & -33 52 51.07 & 13.40 & 12.04 & 11.17 & 10.43 & 10.56 & 10.22 & -- & -- & -- & -- & -- & Class II \\
75 & 17 31 26.86 & -33 54 21.49 & -- & 13.52 & 12.36 & 11.57 & 11.49 & 11.19 & 10.67 & -- & Class II & -- & Class II & --\\
76 & 17 31 27.04 & -33 54 00.50 & -- & -- & -- & 12.69 & 12.72 & 12.22 & 10.47 & -- & -- & -- & Class II & --\\
77 & 17 31 27.43 & -33 53 22.86 & 14.27 & 13.36 & 12.76 & 12.02 &  & 12.09 & 11.81 & -- & -- & -- & -- & Class II \\
78 & 17 31 27.87 & -33 50 34.62 & -- & -- & -- & 11.75 & 11.25 & 10.67 & 9.58 & 6.11 & Class I/II & Class I/II & Class II & --\\
79 & 17 31 28.41 & -33 51 32.64 & 15.32 & 13.33 & 12.22 & 10.82 & 10.59 & 10.30 & 10.33 & -- & -- & -- & -- & Class II \\
80 & 17 31 28.43 & -33 54 15.88 & 12.30 & 10.16 & 9.18 & 8.43 & 8.36 & 7.99 & 7.91 & 5.33 & -- & Class II & -- & -- \\
81 & 17 31 28.57 & -33 50 36.28 & 13.85 & 12.94 & 12.37 & 11.86 & 11.87 & 11.89 & -- & -- & -- & -- & -- & Class II \\
82 & 17 31 28.80 & -33 56 10.96 & 12.59 & 11.68 & 11.04 & 10.57 & 10.61 & 10.45 & 10.46 & -- & -- & -- & -- & Class II \\
83 & 17 31 28.98 & -33 51 28.01 & -- & -- & -- & 13.16 & 13.08 & -- & -- & 5.81 & -- & Class I & -- & --\\
84 & 17 31 30.39 & -33 54 32.58 & 13.33 & 12.41 & 11.88 & -- & -- & -- & -- & -- & -- & -- & -- & Class II \\
85 & 17 31 31.03 & -33 50 16.12 & 15.13 & 12.72 & 11.34 & 10.27 & 10.18 & 9.81 & 9.96 & -- & -- & -- & -- & Class II \\
86 & 17 31 31.15 & -33 54 20.38 & -- & 13.22 & 12.41 & 11.79 & 11.89 & 11.19 & 11.18 & -- & -- & -- & Class II & --\\
87 & 17 31 31.27 & -33 55 28.96 & 12.52 & 11.82 & 11.34 & 10.83 & 10.58 & 10.33 & 10.07 & -- & -- & -- & Class II & --\\
88 & 17 31 32.79 & -33 54 25.12 & 13.80 & 12.88 & 12.31 & -- & -- & -- & -- & -- & -- & -- & -- & Class II \\
89 & 17 31 33.47 & -33 54 10.51 & 11.38 & 10.26 & 9.57 & -- & -- & -- & -- & -- & -- & -- & -- & Class II \\ 
90 & 17 31 34.96 & -33 52 28.45 & 13.28 & 12.19 & 11.76 & 11.46 & 11.43 & 11.29 & 10.89 & -- & Class II & -- & -- & --\\
91 & 17 31 35.17 & -33 50 41.06 & -- & -- & -- & 12.38 & 12.44 & 12.53 & 10.60 & -- & -- & -- & Class II & --\\
92 & 17 31 35.71 & -33 52 50.74 & -- & 13.03 & 11.01 & 9.27 & 9.18 & 9.09 & 8.63 & -- & Class II & -- & -- & --\\
93 & 17 31 37.18 & -33 52 37.60 & -- & -- & -- & 12.41 & 12.38 & 11.95 & 11.39 & -- & Class II & -- & Class II & --\\
\hline 
\end{tabular}
\end{table*}

\end{appendix}

\end{document}